\documentclass[aps, groupedaddress, longbibliography, reprint, superscriptaddress, twocolumn]{revtex4-2}
\usepackage[utf8]{inputenc}
\usepackage{physics}
\usepackage{graphicx}
\usepackage{amsmath}
\usepackage{amssymb}
\usepackage{amssymb}
\usepackage{stmaryrd}
\usepackage{float}
\usepackage{xcolor}
\usepackage[margin=2.5cm]{geometry}
\usepackage[colorlinks,bookmarks=false,citecolor=red,linkcolor=red,urlcolor=blue]{hyperref}
\usepackage{textpos}
\usepackage{natbib}

\begin{document}

\title{Frustration on a centred pyrochlore lattice in metal-organic frameworks}

\author{Rajah Nutakki}
\affiliation{Arnold Sommerfeld Center for Theoretical Physics, University of Munich, Theresienstr. 37, 80333 M\"unchen, Germany}
\affiliation{Munich Center for Quantum Science and Technology (MCQST), Schellingstr. 4, 80799 M\"unchen, Germany}

\author{Richard R\"o\ss-Ohlenroth}
\affiliation{Chair of Solid State and Materials Chemistry, Institute of Physics, University of Augsburg, D-86159 Augsburg, Germany}

\author{Dirk Volkmer}
\affiliation{Chair of Solid State and Materials Chemistry, Institute of Physics, University of Augsburg, D-86159 Augsburg, Germany}

\author{Anton Jesche}
\affiliation{Experimental Physics VI, Center for Electronic Correlations and Magnetism, Institute of Physics, University of Augsburg, 86159 Augsburg, Germany}

\author{Hans-Albrecht Krug von Nidda}
\affiliation{Experimental Physics V, Center for Electronic Correlations and Magnetism, Institute of Physics, University of Augsburg, 86159 Augsburg, Germany}

\author{Alexander A. Tsirlin}
\affiliation{Experimental Physics VI, Center for Electronic Correlations and Magnetism, Institute of Physics, University of Augsburg, 86159 Augsburg, Germany}

\author{Philipp Gegenwart}
\affiliation{Experimental Physics VI, Center for Electronic Correlations and Magnetism, Institute of Physics, University of Augsburg, 86159 Augsburg, Germany}

\author{Lode Pollet}
\affiliation{Arnold Sommerfeld Center for Theoretical Physics, University of Munich, Theresienstr. 37, 80333 M\"unchen, Germany}
\affiliation{Munich Center for Quantum Science and Technology (MCQST), Schellingstr. 4, 80799 M\"unchen, Germany}

\author{Ludovic D.\ C.\ Jaubert}
\affiliation{CNRS, Universit\'e de Bordeaux, LOMA, UMR 5798, 33400 Talence, France}

\date{\today} 
\begin{abstract}
Geometric frustration inhibits magnetic systems from ordering, opening a window to unconventional phases of matter. 
The paradigmatic frustrated lattice in three dimensions to host a spin liquid is the pyrochlore, although there remain few experimental compounds thought to realize such a state. 
Here we go beyond the pyrochlore via molecular design in the metal-azolate framework [Mn(II)(ta)$_2$], which realizes a closely related centred pyrochlore lattice of Mn-spins with $S=5/2$. Despite a Curie-Weiss temperature of $-21$ K indicating the energy scale of magnetic interactions, [Mn(II)(ta)$_2$] orders at only  430 mK, putting it firmly in the category of highly frustrated magnets. Comparing magnetization and specific heat measurements to numerical results for a minimal Heisenberg model, we predict that this material displays distinct features of a classical spin liquid with a structure factor reflecting Coulomb physics in the presence of charges. 
\end{abstract}
\maketitle
Over the last few decades, the theoretical study of magnetism on geometrically frustrated lattices has proven highly successful in identifying exotic states of matter, ranging from classical spin ice \cite{castelnovo2012,spinicebook} to quantum spin liquids \cite{savary2017,zhou2017,knolle2019}.
In 3 dimensions, the most well-studied frustrated lattice is arguably the pyrochlore lattice of corner-sharing tetrahedra.
For the classical nearest neighbor Ising and Heisenberg models on the pyrochlore lattice, the spin liquid ground states have an elegant description in terms of an emergent $U(1)$ gauge field which leads to their characterization as Coulomb spin liquids \cite{isakov2004,henley2010}, with excitations interacting via an effective Coulomb potential and characteristic pinch point singularities in the spin structure factor.
The clearest experimental realization is found in the spin ice compounds $\mathrm{Dy_2Ti_2O_7}$ and $\mathrm{Ho_2Ti_2O_7}$ \cite{bramwell2020}, where the local orientations of the spins and dipolar interactions introduce additional energetic (rather than purely entropic) Coulomb interactions, which in turn leads to a description of excitations in terms of magnetic monopoles \cite{castelnovo2008}.
In the Heisenberg case, recent experiments on the transition metal pyrochlore fluoride, $\mathrm{NaCaNi_2F_7}$, find that it is well described by an $S = 1$ pyrochlore Heisenberg antiferromagnet (PHAF) and thought to realize a Coulomb-like phase~\cite{plumb2019}.
Unfortunately, it is a recurrent theme for Heisenberg materials that further neighbor interactions, anisotropic exchanges or disorder perturbs the spin-liquid physics at low temperatures~\cite{spinicebook,Hallas18,Rau19,trebst22a,Kermarrec21}.

\begin{figure}[t]
	\centering
	\includegraphics[width=8cm]{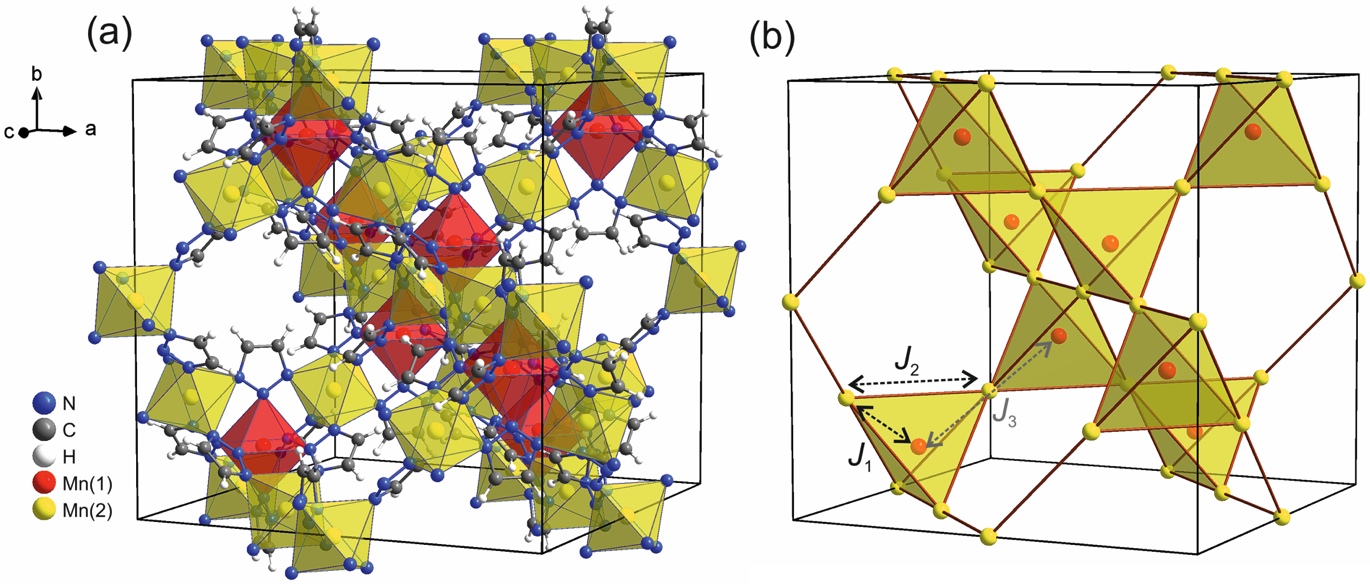}
	\caption{\small{
			\textbf{a,} Combined ball-and-stick and polyhedra model of cubic [Mn(II)(ta)$_2$] highlighting the positions of divalent octahedrally coordinated Mn ions arranged in a diamond-type lattice (CSD code: HEJQEV). The metal centres differ with respect to their special crystallographic positions and coordination environments: Mn(1) is located on Wyckoff position 8b (site symmetry $\bar{4}3m$), coordinated exclusively by the N2 donor atoms of the $\mu_3$-bridging triazolate linker; Mn(2) is found at Wyckoff position 16d (site symmetry $\bar{3}m$), coordinated exclusively by N1 or N3 donor atoms. \textbf{b,} Schematic representation of the centred pyrochlore lattice with magnetic exchange paths for first ($J_{1}$, Mn(1)$-$Mn(2), 3.929\AA), second ($J_{2}$, Mn(2)$-$Mn(2), 6.416\AA) and third ($J_{3}$, Mn(1)$-$Mn(1), 7.858\AA) neighbors. Mn(1) and Mn(2) are respectively labeled centre (orange) and corner (yellow) spins.
	}}
	\label{fig:1}
\end{figure}

%
Recently, metal-organic frameworks have emerged as a class of materials for realizing strongly magnetically frustrated systems \cite{bulled2022}, offering a new avenue to realizing both familiar and novel geometrically frustrated lattices in the lab.
Here, our focus is on the metal-azolate frameworks [M(II)(ta)$_2$], where M(II) is a divalent metal ion and H-ta = 1H-1,2,3-triazole, which exhibit a diamond net with vertices made of M-centred (MM$_4$) tetrahedra [Fig.~\ref{fig:1}], offering the exciting possibility to engineer novel magnetic structures by inserting additional, magnetically active ions into the pyrochlore lattice.
However, so far only a handful of works have addressed their magnetic properties \cite{zhou09,gandara12,grzywa12,Park18,grzywa20,park21}.\\
In this Research Letter, we explore the potential of metal-azolate frameworks in the context of frustrated magnetism by studying [Mn(ta)$_2$], which realizes a centred pyrochlore lattice.
Using a combination of specific-heat and magnetic measurements, ab-initio calculations, Monte Carlo simulations, exact diagonalization and analytical insights, we predict the existence of a finite temperature regime around $T \approx 2 \: \mathrm{K}$ where we expect to find the hallmarks of an underlying classical spin liquid in the {\it minimal} model.
The correlations in this spin liquid can be understood  in the Coulomb framework as a dense fluid of charges created by the center spins, reminiscent of the monopole fluid studied in the context of spin ice \cite{borzi2013,slobinsky2018}.\\
\textbf{Minimal model} -- 
Adapting the synthesis procedure of Ref.~[\onlinecite{He17}], [Mn(ta)$_2$] was prepared as a white powder sample. Rietveld refinements of X-ray diffraction data find that [Mn(ta)$_2$] has the cubic symmetry of the $Fd\bar{3}m$ space group \cite{gandara12} which we confirm through high-resolution X-ray powder diffraction at 5 K. Since the 3d valence band of high-spin Mn(II) ions is half filled, we expect its magnetic moment to be isotropic. Previous conductivity measurements together with periodic DFT band structure calculations \cite{Sun17} classified this compound as a wide-bandgap semiconductor with a bandgap $\Delta=3.1~\textrm{eV} \equiv 36\, 000~\textrm{K}$. We therefore consider [Mn(ta)$_2$] an insulator at temperatures $T\ll \Delta$.
To assess the relevance of different exchange pathways [Fig.~\ref{fig:1}], we performed DFT calculations assuming isotropic exchanges for high-spin 3d$^5$ Mn(II) ions between first $(J_{1})$, second $(J_{2})$ and third $(J_{3})$ neighbors [see Fig.~\ref{fig:1}.b and Supplementary Information~\cite{supplemental} ]. 
We obtain that $J_1^{\rm DFT}\sim 2-4$~K and $\gamma^{\rm DFT}\equiv J_1^{\rm DFT}/J_2^{\rm DFT}\approx 1.3 - 1.65$. The fact that $J_1^{\rm DFT}$ and $J_2^{\rm DFT}$ are of the same order of magnitude is likely due to the similar exchange pathways, traversing either two or three nitrogen ions respectively along the triazolate ligand. 
On the other hand, no such pathway is available beyond second neighbors, hence $|J_3^{\rm DFT}|<0.01~\mathrm{K}~\ll J_1^{\rm DFT}, J_2^{\rm DFT}$. 
Thanks to this separation of scales, it suffices to define a minimal model (CPy) on the centred pyrochlore lattice with only first and second neighbor isotropic couplings, 
\begin{equation}
H = J_1 \sum_{\langle ij \rangle} \mathbf{S}_i \cdot \mathbf{S}_j + 
J_2 \sum_{\langle \langle ij \rangle \rangle} \mathbf{S}_i \cdot \mathbf{S}_j 
- \mu_{0}\mathbf{H} \cdot \sum_i \mathbf{S}_i.
\label{eq:ham}
\end{equation}
%
Confirming that [Mn(ta)$_{2}$] is described by Hamiltonian (\ref{eq:ham}), requires the appropriate theoretical tools. 
Fortunately, owing to the large magnetic moment $S=5/2$ we can resort to classical Monte Carlo simulations, which  have proven to be powerful techniques to describe magnetic properties of pyrochlore materials capable of capturing  long-range order as well as  unconventional correlations of spin liquids \cite{spinicebook,Javanparast15,Yan17,Zhang19,Samarakoon20,Yahne21a}. Continuous-spin models cannot, however, reproduce specific heat at very low temperatures because entropy is ill defined. There, one needs to consider the discrete nature of quantum spins via {\it e.g.} exact diagonalization (ED). Since ED for $S=5/2$ is particularly costly in computer time and memory, we will restrict calculations to a single tetrahdral unit of five spins, fitted to high-temperature experimental data. Such an approximation shall provide an independent estimate of the energy scales, to be compared with parameters obtained from DFT and Monte Carlo simulations. 
In the rest of this paper, $\mathbf{S}_i$ will denote a 3-component classical spin of length $|\mathbf{S}_{i}|=5/2$, except when analyzing specific-heat data where $\mathbf{S}_i$ is a quantum $S=5/2$ spin.\\
\begin{figure*}[t]
\includegraphics[width=\textwidth]{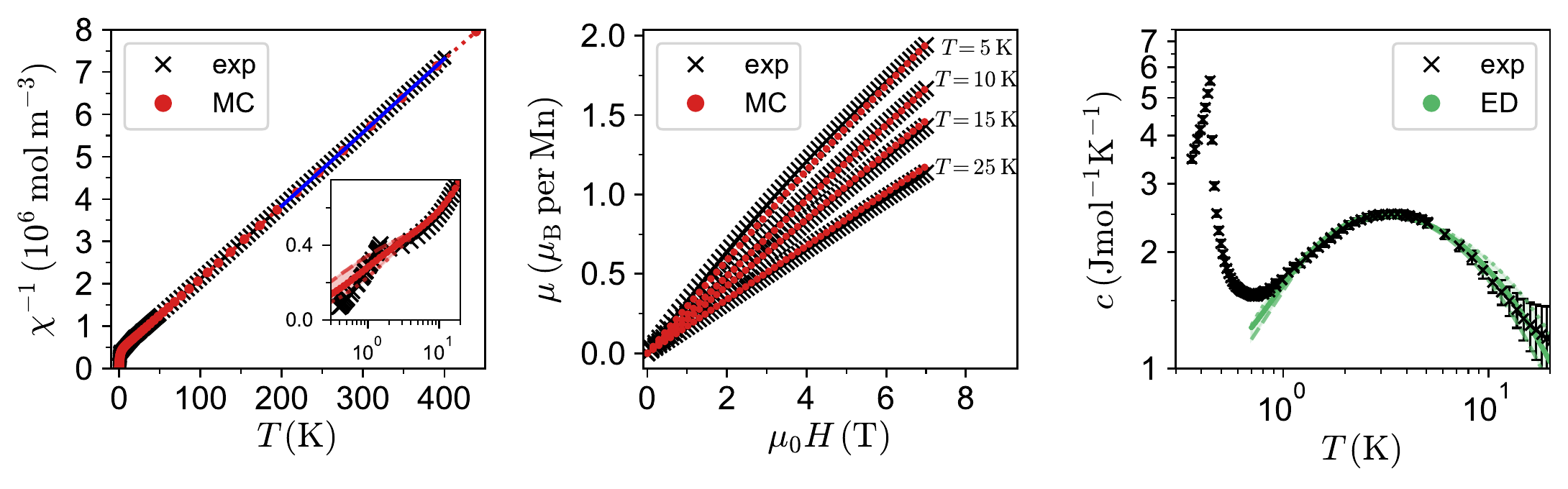}
\caption{\small{
\textbf{Comparison between experiment and theory for [Mn(ta)$_{2}$].}
\textbf{a.} Curie-Weiss fit (blue, for $T > 200~\mathrm{K}$) of the inverse susceptibility $\chi^{-1}$ obtained from magnetization measurements $\chi=M/H$ (black) and MC simulations of over $41 000$ spins (red). Inset, the inverse susceptibilities at low temperatures.
The Curie-Weiss fit yields a Land\'{e} factor of $g = 2.05$.
\textbf{b.} Magnetization in an external field over a broad temperature range.
MC simulations compare well to experimental magnetization measurements at $T > 1\: \mathrm{K}$ for $J_1^{\mathrm MC} = 2.0\: \mathrm{K}$, $\gamma^{\rm MC}=1.51\pm 0.15$.
\textbf{c.} The specific heat of [Mn(ta)$_{2}$] displays a broad bump at $\sim 4$~K and a sharp peak at 0.43 K. The former is qualitatively reproduced by full diagonalization for $J_1^{\rm ED} = 1.95$~K and $\gamma^{\rm ED}\sim 1.75$ up to a rescaling by a factor of 0.8 along the $y-$axis.
In order to show the $\gamma$ dependence of Hamiltonian (\ref{eq:ham}), simulation curves in \textbf{a.} and \textbf{c.} are supplemented by dotted and dashed lines corresponding to $\pm 10\%$ variations of $\gamma$.
}}
\begin{textblock}{1}(-0.2,-5)
	\textbf{a.}
\end{textblock}
\begin{textblock}{1}(3.8,-5)
	\textbf{b.}
\end{textblock}
\begin{textblock}{1}(7.9,-5)
	\textbf{c.}
\end{textblock}
\label{fig:2}
\end{figure*}
\textbf{Comparison between experiment and theory -- }
To explore the magnetism of [Mn(ta)$_2$], we measured magnetic susceptibility over three decades in temperature. 
Fitting the high-temperature regime ($T > 200$ K) of the susceptibility $\chi$ with a Curie-Weiss law [Fig.~\ref{fig:2}.a], we obtain an effective magnetic moment $\mu_{\text{eff}} = 6.05~\mu_B$ per Mn ion and a Curie-Weiss temperature $\Theta_{CW}= -21$~K. 
These values agree with a previous report \cite{gandara12} ($5.8~\mu_B$ and $-21.9$~K respectively) and with the expected size of the magnetic moment $g_{S}\sqrt{S(S+1)}=5.9~\mu_{B}$ which is a stable value for a Mn(II) ion with high-spin 3d$^5$ electronic configuration. $\Theta_{CW}$ indicates sizeable antiferromagnetic interactions. 
However, our specific heat measurements do not find a transition until  a low $T_c = 0.43$~K  [Fig.~\ref{fig:2}.c].
For comparison, the degree of frustration $f\equiv\frac{|\Theta_{CW}|}{T_c} = 49$ is of the same order as the one of the celebrated Kitaev materials \cite{winter17a,trebst22a} and substantially larger than the one of most rare-earth pyrochlore oxides \cite{spinicebook,Hallas18,Rau19}. It means that not only does [Mn(ta)$_{2}$]  offer an unexplored geometry, but one can also expect a sizeable temperature range above $T_{c}$ where frustration is  important.\\
Our main result is the agreement between experiments and Monte Carlo simulations for the (i) magnetic susceptibility in Fig.~\ref{fig:2}.a, and (ii) magnetization curves in Fig.~\ref{fig:2}.b, using coupling parameters that confirm the DFT estimates, $J_1^{\rm MC} = 2.0$~K and $\gamma^{\rm MC}=1.51\pm 0.15$. Further support for this claim is seen in the fit of the specific heat  [Fig.~\ref{fig:2}.c]:
The bump at $\sim 4$~K is also consistent with finite-temperature exact diagonalization for $J_1^{\rm ED} = 1.95$~K and $\gamma^{\rm ED}\sim 1.75$ [Fig.~\ref{fig:2}.c]. 
While the value of $\gamma^{\rm ED}$ should only be considered as a qualitative estimate, the ED results indicate that the bump at 4 K corresponds to a growth of the correlation length beyond a single frustrated unit.
The region $0.43 < T \lesssim 4$~K thus appears appropriate to support a potentially exotic magnetic structure, where frustrated correlations exist, but have not yet been destroyed by long-range order. We study the nature of this regime in the following.\\
\begin{figure*}
	\includegraphics[width=\textwidth]{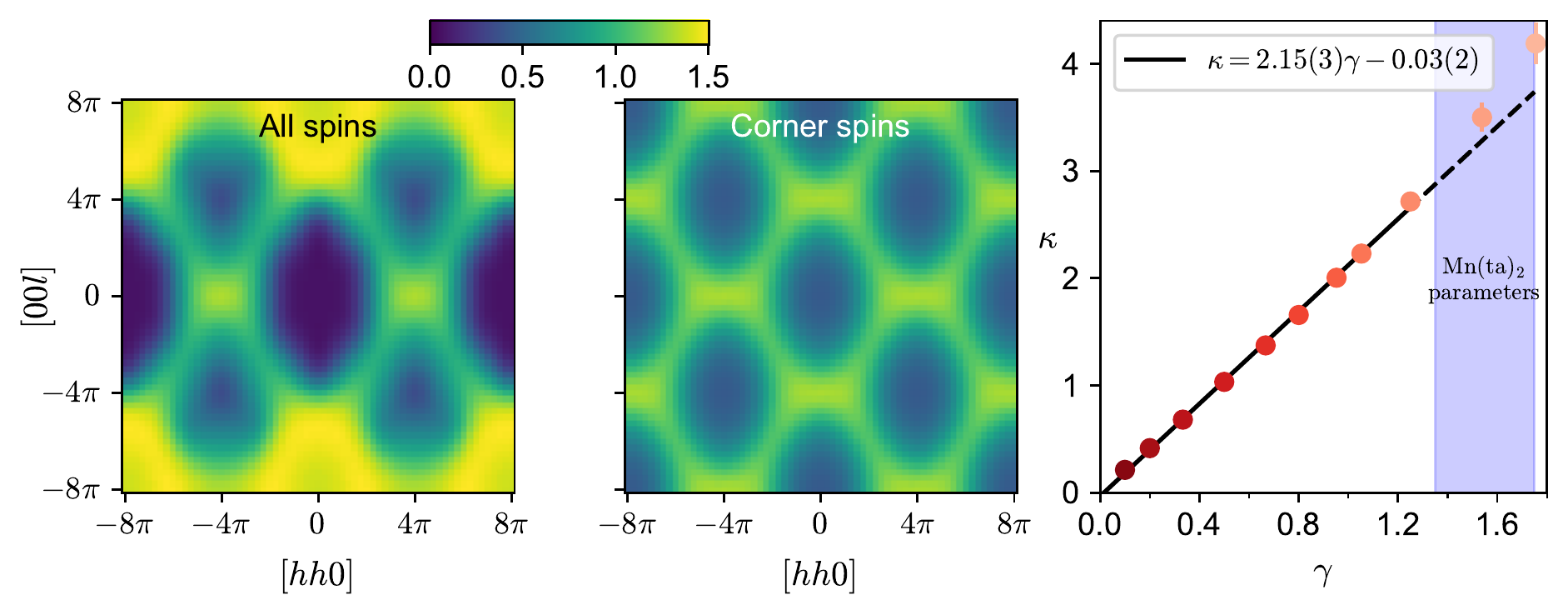}
	\caption{\small{
			Spin structure factor from large-scale MC simulations at $T = 1.5 \: \mathrm{K}, \gamma = 1.5$ in the $[hhl]$ plane including all spins (\textbf{a.}) and only sites residing at the corners of tetrahedra (\textbf{b.}). Finite width bow ties at $\mathbf{q} = (\pm 4\pi, \pm 4\pi, 0)$ are found in both and the structure factor in (\textbf{b.}). is within $1\%$ of that found at $T = 0.07 \: \mathrm{K}$. \textbf{c.} The width of the bow ties in the structure factor of the effective field parametrized by $\kappa$, see Eq. \ref{eq:B}, for different values of the coupling ratio $\gamma$ at $T = 0.07 \: \mathrm{K}$. A linear fit was performed to the data up to $\gamma = 1.25$ (black line).
	}}
	\begin{textblock}{1}(-0.3,-4.4)
		\textbf{a.}
	\end{textblock}
	\begin{textblock}{1}(4,-4.4)
		\textbf{b.}
	\end{textblock}
	\begin{textblock}{1}(7.65,-4.9)
		\textbf{c.}
	\end{textblock}
	\label{fig:3}
\end{figure*}
\textbf{The Centred Pyrochlore Spin Liquid} --
To show how the CPy model can host a spin liquid, it is convenient to rewrite Hamiltonian (\ref{eq:ham}) as a sum over (centred) tetrahedra, with index $\alpha$,
\begin{eqnarray}
H = \frac{J_{2}}{2} \sum_\alpha \abs{\mathbf{L}_\alpha}^2 + \mathrm{const},
\label{eq:HL}\\
\textrm{with}\;\mathbf{L}_\alpha = \gamma \mathbf{S}_{\alpha,c} + \sum_{m =1}^{4} \mathbf{S}_{\alpha,m},
\label{eq:local_constraint}
\end{eqnarray}
where $c$ labels the centre spin while the sum over $m$ runs over the four corner spins. 
The ground state manifold is therefore defined by minimizing $L_\alpha = \abs{\mathbf{L}_\alpha}$ on all units. 
For $\gamma \geq 4$, this yields a long-range ordered ferrimagnet with $1/3$ saturated magnetization where $\mathbf{S}_{\alpha,i=1,..,4}=-\mathbf{S}_{\alpha,c}$. 
On the other hand, for $\gamma < 4$, the ground state is defined by the local constraint
\begin{equation}
\mathbf{L}_\alpha=0, \; \forall \alpha.
\label{eq:constraint}
\end{equation}
Such an energetic constraint, familiar from the kagome and pyrochlore lattices, is often used to define a classical spin liquid, provided order by disorder does not select an ordered state.
A Maxwellian counting argument \cite{Moessner98_geo} gives the degrees of freedom $D_{n}$ of these spin liquids.
For the kagome, $D_{3}=0$ (which is famously marginally disordered), whereas $D_{4}=N_{u}$ for the pyrochlore \cite{Moessner98_geo}, where $N_u$ is the number of units making up the lattice. 
In the CPy model, we find $D_{5}=3\,N_{u}$ for $\gamma\sim 1$.
Similarly, these spin liquids manifest themselves via a number of flat bands $F_{n}$ as the ground state of their excitation spectrum \cite{Reimers91}: $F_{3}=1$ out of 3 bands for kagome, $F_{4}=2$ out of 4 for pyrochlore, whereas
$F_{5}=4$ flat bands out of 6 for $\gamma <4$ in the CPy model (see supplementary information).
Therefore, the CPy model hints at a notably strong spin liquid, even more disordered than the extensively studied ones on kagome and pyrochlore.
This is supported by our Monte Carlo simulations~\cite{Nutakki2023} showing that CPy lies in the middle of a  disordered ground state (found for any $\gamma \lesssim 3$) which indicates that the ordering mechanism in [Mn(ta)$_2$] lies beyond our minimal CPy model. 
This is often the case in frustrated magnetism \cite{spinicebook,Hallas18,Rau19,trebst22a}, where perturbations ultimately lift the ground-state degeneracy in materials. 
In this case, the largest perturbation is likely dipolar interactions with a nearest neighbour strength of $270 \: \mathrm{mK}$. 
Adding these to the CPy model, we find ordering at $250 \mathrm{mK}$ in MC simulations, where the ordered state is an unsaturated ferrimagnet with corner spins realizing a planar antiferromagnet on each tetrahedron with the remaining spin weight. 
Further details are provided in the supplementary information.
Our simulations show that the addition of dipolar interactions does not significantly modify the properties of the model in the regime $1 < T < 4 \: \mathrm{K}$, where we expect the spin liquid to persist.

\textbf{An emergent charge fluid} -- 
To better understand the nature of this spin liquid and its relation to known pyrochlore physics we look at the magnetic correlations, best visualised through the structure factor in reciprocal space, $\mathcal{S}(\mathbf{q})$ [Figs.~\ref{fig:3}.a,b]. 
We find that, in the regime $0 < \gamma < 2$, $\mathcal{S}(\mathbf{q})$ is characterized by finite width bow ties, which are found to arise from the correlations between spins residing at the vertices of the tetrahedra.
The magnetic structure evolves continuously in this regime, with the width of the bow ties increasing with $\gamma$. 
This suggests that these correlations can be understood in the framework of the Coulomb phase on the pyrochlore lattice.
Rewriting the constraint, Eq.~(\ref{eq:constraint}), for each of the spin components $a\in\{x,y,z\}$ as
\begin{equation}
	\sum_{m =1}^{4} S^{a}_{\alpha,m}= - \gamma S^{a}_{\alpha,c}=\rho^{a}_{\alpha}
	\label{eq:constraintmono}
\end{equation}
we interpret $\rho^{a}_{\alpha}$ as a pseudoscalar (``magnetic'') charge, with a strength parameterized by $\gamma$.
Hence, the Coulomb description is that of an effective field coupled to charges on the diamond lattice that break the zero-divergence constraint.
For small $\gamma$, the charge concentration is low, so Debye-Hueckel theory \cite{levin2002,castelnovo2011} can be used to understand the spin correlations.
In this regime, there will be entropic screening of the effective field, resulting in a Lorentzian form for its structure factor
\begin{equation}
	\mathcal{B}^a_{aa}(q_a, q_b = 0, q_c =0) \propto \frac{1}{q_a^2 + \kappa^2},
	\label{eq:B}
\end{equation}
where $a,b,c \in \{x,y,z \}$ and Debye-Hueckel theory predicts that $\kappa \propto \gamma$.
At some value of $\gamma$, the charge concentration becomes large and we will need to account for additional corrections.
Remarkably, we find that the prediction from Debye-Hueckel theory holds in MC simulations up to $\gamma \approx 1.25$, see Figs.~\ref{fig:3}.c.
Thus, the correct description for the regime relevant to [Mn(ta)$_2$], $\gamma^{\mathrm{MC}} \approx 1.5$, is that of a moderately dense charge fluid, which is the Heisenberg model variant of a monopole fluid in spin ice.
Whilst this description accounts for the entropic selection of specific spin configurations, it does not account for energetic considerations necessary for a full effective description of the structure factor, such as which charge distributions enter the ground state.\\
\textbf{Outlook} -- 
The CPy model, established as a minimal model for [Mn(ta)$_2$],  displays a number of attractive features such as large, isotropic Heisenberg interactions, an unusually large number of magnetically disordered degrees of freedom (in fact, higher than for most known spin liquids), a reasonably large Ramirez frustration ratio, and a structure factor reflecting Coulomb physics in the presence of charges controlled through $\gamma$. 
Experimentally, external pressure may be a viable tool for controlling $\gamma$, similar to spin-1/2 frustrated magnets ~\cite{Kermarrec2017,Zayed2017}.
The consequences of this picture for the inelastic spectrum and the nature of excitations are interesting  open questions.
Inspired by routes taken for the pyrochlores~\cite{Ross11,Yan17,Rau19}, ways to build a multitude of exotic phases by taking a selection of degrees of freedom out of the ground state, {\it e.g.},  via chemical substitution or with magnetic field, can be thought of.
Even more broadly, [Mn(ta)$_2$] belongs to a family of metal-azolate frameworks, whose potential for low energy physics remained till recently to a large extent  uncharted. 
These frameworks provide a versatile platform to engineer (quantum) frustrated magnetism on the centred pyrochlore lattice and beyond, such as [Fe(ta)$_2$(BF$_{4}$)$_{x}$] with a degree of frustration $f \approx 27$ \cite{Park18}, [Cu(ta)$_2$] with Cu(II) dimers at low temperature \cite{grzywa12}, and [Cr(II/III)(ta)$_2$(CF$_3$SO$_3$)$_{0.33}$] with large exchange couplings~\cite{park21}. While ligand substitution in similar 1,2,3-triazolate-based Mn(II) networks is known to influence their magnetic properties~\cite{RossOhlenroth2022}, guest molecule loading in the framework pores should further potentiate the adjustment possibilities in such materials. Thus, they offer us a near infinite playground for design and experimental characterization.\\

\begin{acknowledgments}
\textbf{Acknowledgments} --  
We are grateful to Nic Shannon for stimulating discussions. D.V. and R.R.-O. are grateful for financial support from DFG grant VO-829/12-2 (DFG Priority Program 1928 ``Coordination Networks: Building Blocks for Functional Systems'') and SEM micrographs taken by Ralph Freund. A.A.T., A.J., P.G., and H.-A. K.v.N. were supported by the German Research Foundation via TRR80. L.D.C.J. acknowledges financial support from the ``Agence Nationale de la Recherche'' under Grant No.ANR-18-CE30-0011-01. R.N., L.P. and L.D.C.J. acknowledge financial support from the LMU-Bordeaux Research Cooperation Program. R. N. and L. P. acknowledge support from FP7/ERC Consolidator Grant QSIMCORR, No. 771891, and the Deutsche Forschungsgemeinschaft (DFG, German Research Foundation) under Germany's Excellence Strategy -- EXC-2111 -- 390814868. Our simulations make use of the ALPSCore library~\cite{ALPSCore}. 
\end{acknowledgments}



\bibliography{refs}

\begin{thebibliography}{42}%
\makeatletter
\providecommand \@ifxundefined [1]{%
 \@ifx{#1\undefined}
}%
\providecommand \@ifnum [1]{%
 \ifnum #1\expandafter \@firstoftwo
 \else \expandafter \@secondoftwo
 \fi
}%
\providecommand \@ifx [1]{%
 \ifx #1\expandafter \@firstoftwo
 \else \expandafter \@secondoftwo
 \fi
}%
\providecommand \natexlab [1]{#1}%
\providecommand \enquote  [1]{``#1''}%
\providecommand \bibnamefont  [1]{#1}%
\providecommand \bibfnamefont [1]{#1}%
\providecommand \citenamefont [1]{#1}%
\providecommand \href@noop [0]{\@secondoftwo}%
\providecommand \href [0]{\begingroup \@sanitize@url \@href}%
\providecommand \@href[1]{\@@startlink{#1}\@@href}%
\providecommand \@@href[1]{\endgroup#1\@@endlink}%
\providecommand \@sanitize@url [0]{\catcode `\\12\catcode `\$12\catcode
  `\&12\catcode `\#12\catcode `\^12\catcode `\_12\catcode `\%12\relax}%
\providecommand \@@startlink[1]{}%
\providecommand \@@endlink[0]{}%
\providecommand \url  [0]{\begingroup\@sanitize@url \@url }%
\providecommand \@url [1]{\endgroup\@href {#1}{\urlprefix }}%
\providecommand \urlprefix  [0]{URL }%
\providecommand \Eprint [0]{\href }%
\providecommand \doibase [0]{https://doi.org/}%
\providecommand \selectlanguage [0]{\@gobble}%
\providecommand \bibinfo  [0]{\@secondoftwo}%
\providecommand \bibfield  [0]{\@secondoftwo}%
\providecommand \translation [1]{[#1]}%
\providecommand \BibitemOpen [0]{}%
\providecommand \bibitemStop [0]{}%
\providecommand \bibitemNoStop [0]{.\EOS\space}%
\providecommand \EOS [0]{\spacefactor3000\relax}%
\providecommand \BibitemShut  [1]{\csname bibitem#1\endcsname}%
\let\auto@bib@innerbib\@empty
\bibitem [{\citenamefont {Castelnovo}\ \emph {et~al.}(2012)\citenamefont
  {Castelnovo}, \citenamefont {Moessner},\ and\ \citenamefont
  {Sondhi}}]{castelnovo2012}%
  \BibitemOpen
  \bibfield  {author} {\bibinfo {author} {\bibfnamefont {C.}~\bibnamefont
  {Castelnovo}}, \bibinfo {author} {\bibfnamefont {R.}~\bibnamefont
  {Moessner}},\ and\ \bibinfo {author} {\bibfnamefont {S.}~\bibnamefont
  {Sondhi}},\ }\bibfield  {title} {\bibinfo {title} {Spin {{Ice}},
  {{Fractionalization}}, and {{Topological Order}}},\ }\href
  {https://doi.org/10.1146/annurev-conmatphys-020911-125058} {\bibfield
  {journal} {\bibinfo  {journal} {Annual Review of Condensed Matter Physics}\
  }\textbf {\bibinfo {volume} {3}},\ \bibinfo {pages} {35} (\bibinfo {year}
  {2012})}\BibitemShut {NoStop}%
\bibitem [{\citenamefont {Jaubert}\ and\ \citenamefont
  {Udagawa}(2021)}]{spinicebook}%
  \BibitemOpen
  \bibinfo {editor} {\bibfnamefont {L.~D.~C.}\ \bibnamefont {Jaubert}}\ and\
  \bibinfo {editor} {\bibfnamefont {M.}~\bibnamefont {Udagawa}},\ eds.,\
  \href@noop {} {\emph {\bibinfo {title} {Spin Ice}}},\ \bibinfo {series}
  {Springer Series in Solid-State Sciences}, Vol.\ \bibinfo {volume} {197}\
  (\bibinfo  {publisher} {Springer},\ \bibinfo {year} {2021})\BibitemShut
  {NoStop}%
\bibitem [{\citenamefont {Savary}\ and\ \citenamefont
  {Balents}(2017)}]{savary2017}%
  \BibitemOpen
  \bibfield  {author} {\bibinfo {author} {\bibfnamefont {L.}~\bibnamefont
  {Savary}}\ and\ \bibinfo {author} {\bibfnamefont {L.}~\bibnamefont
  {Balents}},\ }\bibfield  {title} {\bibinfo {title} {Quantum {{Spin
  Liquids}}},\ }\href {https://doi.org/10.1088/0034-4885/80/1/016502}
  {\bibfield  {journal} {\bibinfo  {journal} {Reports on Progress in Physics}\
  }\textbf {\bibinfo {volume} {80}},\ \bibinfo {pages} {016502} (\bibinfo
  {year} {2017})},\ \Eprint {https://arxiv.org/abs/1601.03742}
  {arXiv:1601.03742} \BibitemShut {NoStop}%
\bibitem [{\citenamefont {Zhou}\ \emph {et~al.}(2017)\citenamefont {Zhou},
  \citenamefont {Kanoda},\ and\ \citenamefont {Ng}}]{zhou2017}%
  \BibitemOpen
  \bibfield  {author} {\bibinfo {author} {\bibfnamefont {Y.}~\bibnamefont
  {Zhou}}, \bibinfo {author} {\bibfnamefont {K.}~\bibnamefont {Kanoda}},\ and\
  \bibinfo {author} {\bibfnamefont {T.-K.}\ \bibnamefont {Ng}},\ }\bibfield
  {title} {\bibinfo {title} {Quantum spin liquid states},\ }\href
  {https://doi.org/10.1103/RevModPhys.89.025003} {\bibfield  {journal}
  {\bibinfo  {journal} {Reviews of Modern Physics}\ }\textbf {\bibinfo {volume}
  {89}},\ \bibinfo {pages} {025003} (\bibinfo {year} {2017})}\BibitemShut
  {NoStop}%
\bibitem [{\citenamefont {Knolle}\ and\ \citenamefont
  {Moessner}(2019)}]{knolle2019}%
  \BibitemOpen
  \bibfield  {author} {\bibinfo {author} {\bibfnamefont {J.}~\bibnamefont
  {Knolle}}\ and\ \bibinfo {author} {\bibfnamefont {R.}~\bibnamefont
  {Moessner}},\ }\bibfield  {title} {\bibinfo {title} {A {{Field Guide}} to
  {{Spin Liquids}}},\ }\href
  {https://doi.org/10.1146/annurev-conmatphys-031218-013401} {\bibfield
  {journal} {\bibinfo  {journal} {Annual Review of Condensed Matter Physics}\
  }\textbf {\bibinfo {volume} {10}},\ \bibinfo {pages} {451} (\bibinfo {year}
  {2019})}\BibitemShut {NoStop}%
\bibitem [{\citenamefont {Isakov}\ \emph {et~al.}(2004)\citenamefont {Isakov},
  \citenamefont {Gregor}, \citenamefont {Moessner},\ and\ \citenamefont
  {Sondhi}}]{isakov2004}%
  \BibitemOpen
  \bibfield  {author} {\bibinfo {author} {\bibfnamefont {S.~V.}\ \bibnamefont
  {Isakov}}, \bibinfo {author} {\bibfnamefont {K.}~\bibnamefont {Gregor}},
  \bibinfo {author} {\bibfnamefont {R.}~\bibnamefont {Moessner}},\ and\
  \bibinfo {author} {\bibfnamefont {S.~L.}\ \bibnamefont {Sondhi}},\ }\bibfield
   {title} {\bibinfo {title} {Dipolar {{Spin Correlations}} in {{Classical
  Pyrochlore Magnets}}},\ }\href
  {https://doi.org/10.1103/PhysRevLett.93.167204} {\bibfield  {journal}
  {\bibinfo  {journal} {Physical Review Letters}\ }\textbf {\bibinfo {volume}
  {93}},\ \bibinfo {pages} {167204} (\bibinfo {year} {2004})}\BibitemShut
  {NoStop}%
\bibitem [{\citenamefont {Henley}(2010)}]{henley2010}%
  \BibitemOpen
  \bibfield  {author} {\bibinfo {author} {\bibfnamefont {C.~L.}\ \bibnamefont
  {Henley}},\ }\bibfield  {title} {\bibinfo {title} {The ``{{Coulomb Phase}}''
  in {{Frustrated Systems}}},\ }\href
  {https://doi.org/10.1146/annurev-conmatphys-070909-104138} {\bibfield
  {journal} {\bibinfo  {journal} {Annual Review of Condensed Matter Physics}\
  }\textbf {\bibinfo {volume} {1}},\ \bibinfo {pages} {179} (\bibinfo {year}
  {2010})}\BibitemShut {NoStop}%
\bibitem [{\citenamefont {Bramwell}\ and\ \citenamefont
  {Harris}(2020)}]{bramwell2020}%
  \BibitemOpen
  \bibfield  {author} {\bibinfo {author} {\bibfnamefont {S.~T.}\ \bibnamefont
  {Bramwell}}\ and\ \bibinfo {author} {\bibfnamefont {M.~J.}\ \bibnamefont
  {Harris}},\ }\bibfield  {title} {\bibinfo {title} {The history of spin ice},\
  }\href {https://doi.org/10.1088/1361-648X/ab8423} {\bibfield  {journal}
  {\bibinfo  {journal} {Journal of Physics: Condensed Matter}\ }\textbf
  {\bibinfo {volume} {32}},\ \bibinfo {pages} {374010} (\bibinfo {year}
  {2020})}\BibitemShut {NoStop}%
\bibitem [{\citenamefont {Castelnovo}\ \emph {et~al.}(2008)\citenamefont
  {Castelnovo}, \citenamefont {Moessner},\ and\ \citenamefont
  {Sondhi}}]{castelnovo2008}%
  \BibitemOpen
  \bibfield  {author} {\bibinfo {author} {\bibfnamefont {C.}~\bibnamefont
  {Castelnovo}}, \bibinfo {author} {\bibfnamefont {R.}~\bibnamefont
  {Moessner}},\ and\ \bibinfo {author} {\bibfnamefont {S.~L.}\ \bibnamefont
  {Sondhi}},\ }\bibfield  {title} {\bibinfo {title} {Magnetic monopoles in spin
  ice},\ }\href {https://doi.org/10.1038/nature06433} {\bibfield  {journal}
  {\bibinfo  {journal} {Nature}\ }\textbf {\bibinfo {volume} {451}},\ \bibinfo
  {pages} {42} (\bibinfo {year} {2008})}\BibitemShut {NoStop}%
\bibitem [{\citenamefont {Plumb}\ \emph {et~al.}(2019)\citenamefont {Plumb},
  \citenamefont {Changlani}, \citenamefont {Scheie}, \citenamefont {Zhang},
  \citenamefont {Krizan}, \citenamefont {{Rodriguez-Rivera}}, \citenamefont
  {Qiu}, \citenamefont {Winn}, \citenamefont {Cava},\ and\ \citenamefont
  {Broholm}}]{plumb2019}%
  \BibitemOpen
  \bibfield  {author} {\bibinfo {author} {\bibfnamefont {K.~W.}\ \bibnamefont
  {Plumb}}, \bibinfo {author} {\bibfnamefont {H.~J.}\ \bibnamefont
  {Changlani}}, \bibinfo {author} {\bibfnamefont {A.}~\bibnamefont {Scheie}},
  \bibinfo {author} {\bibfnamefont {S.}~\bibnamefont {Zhang}}, \bibinfo
  {author} {\bibfnamefont {J.~W.}\ \bibnamefont {Krizan}}, \bibinfo {author}
  {\bibfnamefont {J.~A.}\ \bibnamefont {{Rodriguez-Rivera}}}, \bibinfo {author}
  {\bibfnamefont {Y.}~\bibnamefont {Qiu}}, \bibinfo {author} {\bibfnamefont
  {B.}~\bibnamefont {Winn}}, \bibinfo {author} {\bibfnamefont {R.~J.}\
  \bibnamefont {Cava}},\ and\ \bibinfo {author} {\bibfnamefont {C.~L.}\
  \bibnamefont {Broholm}},\ }\bibfield  {title} {\bibinfo {title} {Continuum of
  quantum fluctuations in a three-dimensional {{S}} = 1 {{Heisenberg}}
  magnet},\ }\href {https://doi.org/10.1038/s41567-018-0317-3} {\bibfield
  {journal} {\bibinfo  {journal} {Nature Physics}\ }\textbf {\bibinfo {volume}
  {15}},\ \bibinfo {pages} {54} (\bibinfo {year} {2019})}\BibitemShut {NoStop}%
\bibitem [{\citenamefont {{Hallas}}\ \emph {et~al.}(2018)\citenamefont
  {{Hallas}}, \citenamefont {{Gaudet}},\ and\ \citenamefont
  {{Gaulin}}}]{Hallas18}%
  \BibitemOpen
  \bibfield  {author} {\bibinfo {author} {\bibfnamefont {A.~M.}\ \bibnamefont
  {{Hallas}}}, \bibinfo {author} {\bibfnamefont {J.}~\bibnamefont {{Gaudet}}},\
  and\ \bibinfo {author} {\bibfnamefont {B.~D.}\ \bibnamefont {{Gaulin}}},\
  }\bibfield  {title} {\bibinfo {title} {{Experimental Insights into
  Ground-State Selection of Quantum XY Pyrochlores}},\ }\href
  {https://doi.org/10.1146/annurev-conmatphys-031016-025218} {\bibfield
  {journal} {\bibinfo  {journal} {Annual Review of Condensed Matter Physics}\
  }\textbf {\bibinfo {volume} {9}},\ \bibinfo {pages} {105} (\bibinfo {year}
  {2018})}\BibitemShut {NoStop}%
\bibitem [{\citenamefont {Rau}\ and\ \citenamefont {Gingras}(2019)}]{Rau19}%
  \BibitemOpen
  \bibfield  {author} {\bibinfo {author} {\bibfnamefont {J.~G.}\ \bibnamefont
  {Rau}}\ and\ \bibinfo {author} {\bibfnamefont {M.~J.}\ \bibnamefont
  {Gingras}},\ }\bibfield  {title} {\bibinfo {title} {Frustrated quantum
  rare-earth pyrochlores},\ }\href
  {https://doi.org/10.1146/annurev-conmatphys-022317-110520} {\bibfield
  {journal} {\bibinfo  {journal} {Annual Review of Condensed Matter Physics}\
  }\textbf {\bibinfo {volume} {10}},\ \bibinfo {pages} {357} (\bibinfo {year}
  {2019})}\BibitemShut {NoStop}%
\bibitem [{\citenamefont {Trebst}\ and\ \citenamefont
  {Hickey}(2022)}]{trebst22a}%
  \BibitemOpen
  \bibfield  {author} {\bibinfo {author} {\bibfnamefont {S.}~\bibnamefont
  {Trebst}}\ and\ \bibinfo {author} {\bibfnamefont {C.}~\bibnamefont
  {Hickey}},\ }\bibfield  {title} {\bibinfo {title} {Kitaev materials},\ }\href
  {https://doi.org/10.1016/j.physrep.2021.11.003} {\bibfield  {journal}
  {\bibinfo  {journal} {Physics Reports}\ }\textbf {\bibinfo {volume} {950}},\
  \bibinfo {pages} {1} (\bibinfo {year} {2022})}\BibitemShut {NoStop}%
\bibitem [{\citenamefont {Kermarrec}\ \emph {et~al.}(2021)\citenamefont
  {Kermarrec}, \citenamefont {Kumar}, \citenamefont {Bernard}, \citenamefont
  {H\'enaff}, \citenamefont {Mendels}, \citenamefont {Bert}, \citenamefont
  {Paulose}, \citenamefont {Hazra},\ and\ \citenamefont
  {Koteswararao}}]{Kermarrec21}%
  \BibitemOpen
  \bibfield  {author} {\bibinfo {author} {\bibfnamefont {E.}~\bibnamefont
  {Kermarrec}}, \bibinfo {author} {\bibfnamefont {R.}~\bibnamefont {Kumar}},
  \bibinfo {author} {\bibfnamefont {G.}~\bibnamefont {Bernard}}, \bibinfo
  {author} {\bibfnamefont {R.}~\bibnamefont {H\'enaff}}, \bibinfo {author}
  {\bibfnamefont {P.}~\bibnamefont {Mendels}}, \bibinfo {author} {\bibfnamefont
  {F.}~\bibnamefont {Bert}}, \bibinfo {author} {\bibfnamefont {P.~L.}\
  \bibnamefont {Paulose}}, \bibinfo {author} {\bibfnamefont {B.~K.}\
  \bibnamefont {Hazra}},\ and\ \bibinfo {author} {\bibfnamefont
  {B.}~\bibnamefont {Koteswararao}},\ }\bibfield  {title} {\bibinfo {title}
  {Classical spin liquid state in the $s=\frac{5}{2}$ heisenberg kagome
  antiferromagnet
  ${\mathrm{li}}_{9}{\mathrm{fe}}_{3}{({\mathrm{P}}_{2}{\mathrm{O}}_{7})}_{3}({\mathrm{po}}_{4}{)}_{2}$},\
  }\href {https://doi.org/10.1103/PhysRevLett.127.157202} {\bibfield  {journal}
  {\bibinfo  {journal} {Phys. Rev. Lett.}\ }\textbf {\bibinfo {volume} {127}},\
  \bibinfo {pages} {157202} (\bibinfo {year} {2021})}\BibitemShut {NoStop}%
\bibitem [{\citenamefont {Bulled}\ \emph {et~al.}(2022)\citenamefont {Bulled},
  \citenamefont {Paddison}, \citenamefont {Wildes}, \citenamefont {Lhotel},
  \citenamefont {Cassidy}, \citenamefont {{Pato-Dold{\'a}n}}, \citenamefont
  {{G{\'o}mez-Aguirre}}, \citenamefont {Saines},\ and\ \citenamefont
  {Goodwin}}]{bulled2022}%
  \BibitemOpen
  \bibfield  {author} {\bibinfo {author} {\bibfnamefont {J.~M.}\ \bibnamefont
  {Bulled}}, \bibinfo {author} {\bibfnamefont {J.~A.~M.}\ \bibnamefont
  {Paddison}}, \bibinfo {author} {\bibfnamefont {A.}~\bibnamefont {Wildes}},
  \bibinfo {author} {\bibfnamefont {E.}~\bibnamefont {Lhotel}}, \bibinfo
  {author} {\bibfnamefont {S.~J.}\ \bibnamefont {Cassidy}}, \bibinfo {author}
  {\bibfnamefont {B.}~\bibnamefont {{Pato-Dold{\'a}n}}}, \bibinfo {author}
  {\bibfnamefont {L.~C.}\ \bibnamefont {{G{\'o}mez-Aguirre}}}, \bibinfo
  {author} {\bibfnamefont {P.~J.}\ \bibnamefont {Saines}},\ and\ \bibinfo
  {author} {\bibfnamefont {A.~L.}\ \bibnamefont {Goodwin}},\ }\bibfield
  {title} {\bibinfo {title} {Geometric {{Frustration}} on the {{Trillium
  Lattice}} in a {{Magnetic Metal-Organic Framework}}},\ }\href
  {https://doi.org/10.1103/PhysRevLett.128.177201} {\bibfield  {journal}
  {\bibinfo  {journal} {Physical Review Letters}\ }\textbf {\bibinfo {volume}
  {128}},\ \bibinfo {pages} {177201} (\bibinfo {year} {2022})}\BibitemShut
  {NoStop}%
\bibitem [{\citenamefont {Zhou}\ \emph {et~al.}(2009)\citenamefont {Zhou},
  \citenamefont {Peng}, \citenamefont {Du}, \citenamefont {Zuo},\ and\
  \citenamefont {You}}]{zhou09}%
  \BibitemOpen
  \bibfield  {author} {\bibinfo {author} {\bibfnamefont {X.-H.}\ \bibnamefont
  {Zhou}}, \bibinfo {author} {\bibfnamefont {Y.-H.}\ \bibnamefont {Peng}},
  \bibinfo {author} {\bibfnamefont {X.-D.}\ \bibnamefont {Du}}, \bibinfo
  {author} {\bibfnamefont {J.-L.}\ \bibnamefont {Zuo}},\ and\ \bibinfo {author}
  {\bibfnamefont {X.-Z.}\ \bibnamefont {You}},\ }\bibfield  {title} {\bibinfo
  {title} {Hydrothermal syntheses and structures of three novel coordination
  polymers assembled from 1,2,3-triazolate ligands},\ }\href
  {https://doi.org/10.1039/b819302a} {\bibfield  {journal} {\bibinfo  {journal}
  {CrystEngComm}\ }\textbf {\bibinfo {volume} {11}},\ \bibinfo {pages} {1964}
  (\bibinfo {year} {2009})}\BibitemShut {NoStop}%
\bibitem [{\citenamefont {Gándara}\ \emph {et~al.}(2012)\citenamefont
  {Gándara}, \citenamefont {Uribe-Romo}, \citenamefont {Britt}, \citenamefont
  {Furukawa}, \citenamefont {Lei}, \citenamefont {Cheng}, \citenamefont {Duan},
  \citenamefont {O'Keeffe},\ and\ \citenamefont {Yaghi}}]{gandara12}%
  \BibitemOpen
  \bibfield  {author} {\bibinfo {author} {\bibfnamefont {F.}~\bibnamefont
  {Gándara}}, \bibinfo {author} {\bibfnamefont {F.~J.}\ \bibnamefont
  {Uribe-Romo}}, \bibinfo {author} {\bibfnamefont {D.~K.}\ \bibnamefont
  {Britt}}, \bibinfo {author} {\bibfnamefont {H.}~\bibnamefont {Furukawa}},
  \bibinfo {author} {\bibfnamefont {L.}~\bibnamefont {Lei}}, \bibinfo {author}
  {\bibfnamefont {R.}~\bibnamefont {Cheng}}, \bibinfo {author} {\bibfnamefont
  {X.}~\bibnamefont {Duan}}, \bibinfo {author} {\bibfnamefont {M.}~\bibnamefont
  {O'Keeffe}},\ and\ \bibinfo {author} {\bibfnamefont {O.~M.}\ \bibnamefont
  {Yaghi}},\ }\bibfield  {title} {\bibinfo {title} {Porous, {Conductive}
  {Metal}-{Triazolates} and {Their} {Structural} {Elucidation} by the
  {Charge}-{Flipping} {Method}},\ }\href
  {https://doi.org/10.1002/chem.201103433} {\bibfield  {journal} {\bibinfo
  {journal} {Chemistry - A European Journal}\ }\textbf {\bibinfo {volume}
  {18}},\ \bibinfo {pages} {10595} (\bibinfo {year} {2012})}\BibitemShut
  {NoStop}%
\bibitem [{\citenamefont {Grzywa}\ \emph {et~al.}(2012)\citenamefont {Grzywa},
  \citenamefont {Denysenko}, \citenamefont {Hanss}, \citenamefont {Scheidt},
  \citenamefont {Scherer}, \citenamefont {Weil},\ and\ \citenamefont
  {Volkmer}}]{grzywa12}%
  \BibitemOpen
  \bibfield  {author} {\bibinfo {author} {\bibfnamefont {M.}~\bibnamefont
  {Grzywa}}, \bibinfo {author} {\bibfnamefont {D.}~\bibnamefont {Denysenko}},
  \bibinfo {author} {\bibfnamefont {J.}~\bibnamefont {Hanss}}, \bibinfo
  {author} {\bibfnamefont {E.-W.}\ \bibnamefont {Scheidt}}, \bibinfo {author}
  {\bibfnamefont {W.}~\bibnamefont {Scherer}}, \bibinfo {author} {\bibfnamefont
  {M.}~\bibnamefont {Weil}},\ and\ \bibinfo {author} {\bibfnamefont
  {D.}~\bibnamefont {Volkmer}},\ }\bibfield  {title} {\bibinfo {title} {{CuN6}
  {Jahn}–{Teller} centers in coordination frameworks comprising fully
  condensed {Kuratowski}-type secondary building units: phase transitions and
  magneto-structural correlations},\ }\href
  {https://doi.org/10.1039/c2dt12311h} {\bibfield  {journal} {\bibinfo
  {journal} {Dalton Transactions}\ }\textbf {\bibinfo {volume} {41}},\ \bibinfo
  {pages} {4239} (\bibinfo {year} {2012})}\BibitemShut {NoStop}%
\bibitem [{\citenamefont {Park}\ \emph {et~al.}(2018)\citenamefont {Park},
  \citenamefont {Aubrey}, \citenamefont {Oktawiec}, \citenamefont {Chakarawet},
  \citenamefont {Darago}, \citenamefont {Grandjean}, \citenamefont {Long},\
  and\ \citenamefont {Long}}]{Park18}%
  \BibitemOpen
  \bibfield  {author} {\bibinfo {author} {\bibfnamefont {J.~G.}\ \bibnamefont
  {Park}}, \bibinfo {author} {\bibfnamefont {M.~L.}\ \bibnamefont {Aubrey}},
  \bibinfo {author} {\bibfnamefont {J.}~\bibnamefont {Oktawiec}}, \bibinfo
  {author} {\bibfnamefont {K.}~\bibnamefont {Chakarawet}}, \bibinfo {author}
  {\bibfnamefont {L.~E.}\ \bibnamefont {Darago}}, \bibinfo {author}
  {\bibfnamefont {F.}~\bibnamefont {Grandjean}}, \bibinfo {author}
  {\bibfnamefont {G.~J.}\ \bibnamefont {Long}},\ and\ \bibinfo {author}
  {\bibfnamefont {J.~R.}\ \bibnamefont {Long}},\ }\bibfield  {title} {\bibinfo
  {title} {Charge {Delocalization} and {Bulk} {Electronic} {Conductivity} in
  the {Mixed}-{Valence} {Metal}–{Organic} {Framework}
  {Fe}(1,2,3-triazolate)2({BF4})x},\ }\href
  {https://doi.org/10.1021/jacs.8b03696} {\bibfield  {journal} {\bibinfo
  {journal} {Journal of the American Chemical Society}\ }\textbf {\bibinfo
  {volume} {140}},\ \bibinfo {pages} {8526} (\bibinfo {year} {2018})},\
  \bibinfo {note} {publisher: American Chemical Society}\BibitemShut {NoStop}%
\bibitem [{\citenamefont {Grzywa}\ \emph {et~al.}(2020)\citenamefont {Grzywa},
  \citenamefont {Röß-Ohlenroth}, \citenamefont {Muschielok}, \citenamefont
  {Oberhofer}, \citenamefont {Błachowski}, \citenamefont {Żukrowski},
  \citenamefont {Vieweg}, \citenamefont {von Nidda},\ and\ \citenamefont
  {Volkmer}}]{grzywa20}%
  \BibitemOpen
  \bibfield  {author} {\bibinfo {author} {\bibfnamefont {M.}~\bibnamefont
  {Grzywa}}, \bibinfo {author} {\bibfnamefont {R.}~\bibnamefont
  {Röß-Ohlenroth}}, \bibinfo {author} {\bibfnamefont {C.}~\bibnamefont
  {Muschielok}}, \bibinfo {author} {\bibfnamefont {H.}~\bibnamefont
  {Oberhofer}}, \bibinfo {author} {\bibfnamefont {A.}~\bibnamefont
  {Błachowski}}, \bibinfo {author} {\bibfnamefont {J.}~\bibnamefont
  {Żukrowski}}, \bibinfo {author} {\bibfnamefont {D.}~\bibnamefont {Vieweg}},
  \bibinfo {author} {\bibfnamefont {H.-A.~K.}\ \bibnamefont {von Nidda}},\ and\
  \bibinfo {author} {\bibfnamefont {D.}~\bibnamefont {Volkmer}},\ }\bibfield
  {title} {\bibinfo {title} {Cooperative {Large}-{Hysteresis}
  {Spin}-{Crossover} {Transition} in the {Iron}({II}) {Triazolate} [{Fe}(ta)2]
  {Metal}–{Organic} {Framework}},\ }\href
  {https://doi.org/10.1021/acs.inorgchem.0c00814} {\bibfield  {journal}
  {\bibinfo  {journal} {Inorganic Chemistry}\ }\textbf {\bibinfo {volume}
  {59}},\ \bibinfo {pages} {10501} (\bibinfo {year} {2020})}\BibitemShut
  {NoStop}%
\bibitem [{\citenamefont {Park}(2021)}]{park21}%
  \BibitemOpen
  \bibfield  {author} {\bibinfo {author} {\bibfnamefont {J.~G.}\ \bibnamefont
  {Park}},\ }\bibfield  {title} {\bibinfo {title} {Magnetic ordering through
  itinerant ferromagnetism in a metal–organic framework},\ }\href@noop {}
  {\bibfield  {journal} {\bibinfo  {journal} {Nature Chemistry}\ }\textbf
  {\bibinfo {volume} {13}},\ \bibinfo {pages} {11} (\bibinfo {year}
  {2021})}\BibitemShut {NoStop}%
\bibitem [{\citenamefont {Borzi}\ \emph {et~al.}(2013)\citenamefont {Borzi},
  \citenamefont {Slobinsky},\ and\ \citenamefont {Grigera}}]{borzi2013}%
  \BibitemOpen
  \bibfield  {author} {\bibinfo {author} {\bibfnamefont {R.~A.}\ \bibnamefont
  {Borzi}}, \bibinfo {author} {\bibfnamefont {D.}~\bibnamefont {Slobinsky}},\
  and\ \bibinfo {author} {\bibfnamefont {S.~A.}\ \bibnamefont {Grigera}},\
  }\bibfield  {title} {\bibinfo {title} {Charge {{Ordering}} in a {{Pure Spin
  Model}}: {{Dipolar Spin Ice}}},\ }\href
  {https://doi.org/10.1103/PhysRevLett.111.147204} {\bibfield  {journal}
  {\bibinfo  {journal} {Physical Review Letters}\ }\textbf {\bibinfo {volume}
  {111}},\ \bibinfo {pages} {147204} (\bibinfo {year} {2013})}\BibitemShut
  {NoStop}%
\bibitem [{\citenamefont {Slobinsky}\ \emph {et~al.}(2018)\citenamefont
  {Slobinsky}, \citenamefont {Baglietto},\ and\ \citenamefont
  {Borzi}}]{slobinsky2018}%
  \BibitemOpen
  \bibfield  {author} {\bibinfo {author} {\bibfnamefont {D.}~\bibnamefont
  {Slobinsky}}, \bibinfo {author} {\bibfnamefont {G.}~\bibnamefont
  {Baglietto}},\ and\ \bibinfo {author} {\bibfnamefont {R.~A.}\ \bibnamefont
  {Borzi}},\ }\bibfield  {title} {\bibinfo {title} {Charge and spin
  correlations in the monopole liquid},\ }\href
  {https://doi.org/10.1103/PhysRevB.97.174422} {\bibfield  {journal} {\bibinfo
  {journal} {Physical Review B}\ }\textbf {\bibinfo {volume} {97}},\ \bibinfo
  {pages} {174422} (\bibinfo {year} {2018})}\BibitemShut {NoStop}%
\bibitem [{\citenamefont {He}\ \emph {et~al.}(2017)\citenamefont {He},
  \citenamefont {Ye}, \citenamefont {Xu}, \citenamefont {Zhou}, \citenamefont
  {Zhou}, \citenamefont {Chen}, \citenamefont {Zhang},\ and\ \citenamefont
  {Chen}}]{He17}%
  \BibitemOpen
  \bibfield  {author} {\bibinfo {author} {\bibfnamefont {C.-T.}\ \bibnamefont
  {He}}, \bibinfo {author} {\bibfnamefont {Z.-M.}\ \bibnamefont {Ye}}, \bibinfo
  {author} {\bibfnamefont {Y.-T.}\ \bibnamefont {Xu}}, \bibinfo {author}
  {\bibfnamefont {D.-D.}\ \bibnamefont {Zhou}}, \bibinfo {author}
  {\bibfnamefont {H.-L.}\ \bibnamefont {Zhou}}, \bibinfo {author}
  {\bibfnamefont {D.}~\bibnamefont {Chen}}, \bibinfo {author} {\bibfnamefont
  {J.-P.}\ \bibnamefont {Zhang}},\ and\ \bibinfo {author} {\bibfnamefont
  {X.-M.}\ \bibnamefont {Chen}},\ }\bibfield  {title} {\bibinfo {title}
  {Hyperfine adjustment of flexible pore-surface pockets enables smart
  recognition of gas size and quadrupole moment},\ }\href
  {https://doi.org/10.1039/C7SC03067C} {\bibfield  {journal} {\bibinfo
  {journal} {Chem. Sci.}\ }\textbf {\bibinfo {volume} {8}},\ \bibinfo {pages}
  {7560} (\bibinfo {year} {2017})}\BibitemShut {NoStop}%
\bibitem [{\citenamefont {Sun}\ \emph {et~al.}(2017)\citenamefont {Sun},
  \citenamefont {Hendon}, \citenamefont {Park}, \citenamefont {Tulchinsky},
  \citenamefont {Wan}, \citenamefont {Wang}, \citenamefont {Walsh},\ and\
  \citenamefont {Dincă}}]{Sun17}%
  \BibitemOpen
  \bibfield  {author} {\bibinfo {author} {\bibfnamefont {L.}~\bibnamefont
  {Sun}}, \bibinfo {author} {\bibfnamefont {C.~H.}\ \bibnamefont {Hendon}},
  \bibinfo {author} {\bibfnamefont {S.~S.}\ \bibnamefont {Park}}, \bibinfo
  {author} {\bibfnamefont {Y.}~\bibnamefont {Tulchinsky}}, \bibinfo {author}
  {\bibfnamefont {R.}~\bibnamefont {Wan}}, \bibinfo {author} {\bibfnamefont
  {F.}~\bibnamefont {Wang}}, \bibinfo {author} {\bibfnamefont {A.}~\bibnamefont
  {Walsh}},\ and\ \bibinfo {author} {\bibfnamefont {M.}~\bibnamefont
  {Dincă}},\ }\bibfield  {title} {\bibinfo {title} {Is iron unique in
  promoting electrical conductivity in {MOFs}?},\ }\href
  {https://doi.org/10.1039/C7SC00647K} {\bibfield  {journal} {\bibinfo
  {journal} {Chemical Science}\ }\textbf {\bibinfo {volume} {8}},\ \bibinfo
  {pages} {4450} (\bibinfo {year} {2017})}\BibitemShut {NoStop}%
\bibitem [{sup()}]{supplemental}%
  \BibitemOpen
  \href@noop {} {\bibinfo {title} {see the supplemental}}\BibitemShut {NoStop}%
\bibitem [{\citenamefont {Javanparast}\ \emph {et~al.}(2015)\citenamefont
  {Javanparast}, \citenamefont {Hao}, \citenamefont {Enjalran},\ and\
  \citenamefont {Gingras}}]{Javanparast15}%
  \BibitemOpen
  \bibfield  {author} {\bibinfo {author} {\bibfnamefont {B.}~\bibnamefont
  {Javanparast}}, \bibinfo {author} {\bibfnamefont {Z.}~\bibnamefont {Hao}},
  \bibinfo {author} {\bibfnamefont {M.}~\bibnamefont {Enjalran}},\ and\
  \bibinfo {author} {\bibfnamefont {M.~J.~P.}\ \bibnamefont {Gingras}},\
  }\bibfield  {title} {\bibinfo {title} {Fluctuation-driven selection at
  criticality in a frustrated magnetic system: The case of
  multiple-$\mathbf{k}$ partial order on the pyrochlore lattice},\ }\href
  {https://doi.org/10.1103/PhysRevLett.114.130601} {\bibfield  {journal}
  {\bibinfo  {journal} {Phys. Rev. Lett.}\ }\textbf {\bibinfo {volume} {114}},\
  \bibinfo {pages} {130601} (\bibinfo {year} {2015})}\BibitemShut {NoStop}%
\bibitem [{\citenamefont {Yan}\ \emph {et~al.}(2017)\citenamefont {Yan},
  \citenamefont {Benton}, \citenamefont {Jaubert},\ and\ \citenamefont
  {Shannon}}]{Yan17}%
  \BibitemOpen
  \bibfield  {author} {\bibinfo {author} {\bibfnamefont {H.}~\bibnamefont
  {Yan}}, \bibinfo {author} {\bibfnamefont {O.}~\bibnamefont {Benton}},
  \bibinfo {author} {\bibfnamefont {L.~D.~C.}\ \bibnamefont {Jaubert}},\ and\
  \bibinfo {author} {\bibfnamefont {N.}~\bibnamefont {Shannon}},\ }\bibfield
  {title} {\bibinfo {title} {Theory of multiple-phase competition in pyrochlore
  magnets with anisotropic exchange with application to
  yb$_{2}$ti$_{2}$o$_{7}$, er$_{2}$ti$_{2}$o$_{7}$, and
  er$_{2}$sn$_{2}$o$_{7}$},\ }\href
  {https://doi.org/10.1103/PhysRevB.95.094422} {\bibfield  {journal} {\bibinfo
  {journal} {Phys. Rev. B}\ }\textbf {\bibinfo {volume} {95}},\ \bibinfo
  {pages} {094422} (\bibinfo {year} {2017})}\BibitemShut {NoStop}%
\bibitem [{\citenamefont {Zhang}\ \emph {et~al.}(2019)\citenamefont {Zhang},
  \citenamefont {Changlani}, \citenamefont {Plumb}, \citenamefont
  {Tchernyshyov},\ and\ \citenamefont {Moessner}}]{Zhang19}%
  \BibitemOpen
  \bibfield  {author} {\bibinfo {author} {\bibfnamefont {S.}~\bibnamefont
  {Zhang}}, \bibinfo {author} {\bibfnamefont {H.~J.}\ \bibnamefont
  {Changlani}}, \bibinfo {author} {\bibfnamefont {K.~W.}\ \bibnamefont
  {Plumb}}, \bibinfo {author} {\bibfnamefont {O.}~\bibnamefont
  {Tchernyshyov}},\ and\ \bibinfo {author} {\bibfnamefont {R.}~\bibnamefont
  {Moessner}},\ }\bibfield  {title} {\bibinfo {title} {Dynamical {Structure}
  {Factor} of the {Three}-{Dimensional} {Quantum} {Spin} {Liquid} {Candidate}
  {NaCaNi} 2 {F} 7},\ }\href {https://doi.org/10.1103/PhysRevLett.122.167203}
  {\bibfield  {journal} {\bibinfo  {journal} {Physical Review Letters}\
  }\textbf {\bibinfo {volume} {122}},\ \bibinfo {pages} {167203} (\bibinfo
  {year} {2019})}\BibitemShut {NoStop}%
\bibitem [{\citenamefont {Samarakoon}\ \emph {et~al.}(2020)\citenamefont
  {Samarakoon}, \citenamefont {Barros}, \citenamefont {Li}, \citenamefont
  {Eisenbach}, \citenamefont {Zhang}, \citenamefont {Ye}, \citenamefont
  {Sharma}, \citenamefont {Dun}, \citenamefont {Zhou}, \citenamefont {Grigera},
  \citenamefont {Batista},\ and\ \citenamefont {Tennant}}]{Samarakoon20}%
  \BibitemOpen
  \bibfield  {author} {\bibinfo {author} {\bibfnamefont {A.~M.}\ \bibnamefont
  {Samarakoon}}, \bibinfo {author} {\bibfnamefont {K.}~\bibnamefont {Barros}},
  \bibinfo {author} {\bibfnamefont {Y.~W.}\ \bibnamefont {Li}}, \bibinfo
  {author} {\bibfnamefont {M.}~\bibnamefont {Eisenbach}}, \bibinfo {author}
  {\bibfnamefont {Q.}~\bibnamefont {Zhang}}, \bibinfo {author} {\bibfnamefont
  {F.}~\bibnamefont {Ye}}, \bibinfo {author} {\bibfnamefont {V.}~\bibnamefont
  {Sharma}}, \bibinfo {author} {\bibfnamefont {Z.~L.}\ \bibnamefont {Dun}},
  \bibinfo {author} {\bibfnamefont {H.}~\bibnamefont {Zhou}}, \bibinfo {author}
  {\bibfnamefont {S.~A.}\ \bibnamefont {Grigera}}, \bibinfo {author}
  {\bibfnamefont {C.~D.}\ \bibnamefont {Batista}},\ and\ \bibinfo {author}
  {\bibfnamefont {D.~A.}\ \bibnamefont {Tennant}},\ }\bibfield  {title}
  {\bibinfo {title} {Machine-learning-assisted insight into spin ice {Dy} 2
  {Ti} 2 {O} 7},\ }\href {https://doi.org/10.1038/s41467-020-14660-y}
  {\bibfield  {journal} {\bibinfo  {journal} {Nature Communications}\ }\textbf
  {\bibinfo {volume} {11}},\ \bibinfo {pages} {892} (\bibinfo {year}
  {2020})}\BibitemShut {NoStop}%
\bibitem [{\citenamefont {Yahne}\ \emph {et~al.}(2021)\citenamefont {Yahne},
  \citenamefont {Pereira}, \citenamefont {Jaubert}, \citenamefont {Sanjeewa},
  \citenamefont {Powell}, \citenamefont {Kolis}, \citenamefont {Xu},
  \citenamefont {Enjalran}, \citenamefont {Gingras},\ and\ \citenamefont
  {Ross}}]{Yahne21a}%
  \BibitemOpen
  \bibfield  {author} {\bibinfo {author} {\bibfnamefont {D.~R.}\ \bibnamefont
  {Yahne}}, \bibinfo {author} {\bibfnamefont {D.}~\bibnamefont {Pereira}},
  \bibinfo {author} {\bibfnamefont {L.~D.~C.}\ \bibnamefont {Jaubert}},
  \bibinfo {author} {\bibfnamefont {L.~D.}\ \bibnamefont {Sanjeewa}}, \bibinfo
  {author} {\bibfnamefont {M.}~\bibnamefont {Powell}}, \bibinfo {author}
  {\bibfnamefont {J.~W.}\ \bibnamefont {Kolis}}, \bibinfo {author}
  {\bibfnamefont {G.}~\bibnamefont {Xu}}, \bibinfo {author} {\bibfnamefont
  {M.}~\bibnamefont {Enjalran}}, \bibinfo {author} {\bibfnamefont {M.~J.~P.}\
  \bibnamefont {Gingras}},\ and\ \bibinfo {author} {\bibfnamefont {K.~A.}\
  \bibnamefont {Ross}},\ }\bibfield  {title} {\bibinfo {title} {Understanding
  reentrance in frustrated magnets: The case of the
  ${\mathrm{er}}_{2}{\mathrm{sn}}_{2}{\mathrm{o}}_{7}$ pyrochlore},\ }\href
  {https://doi.org/10.1103/PhysRevLett.127.277206} {\bibfield  {journal}
  {\bibinfo  {journal} {Phys. Rev. Lett.}\ }\textbf {\bibinfo {volume} {127}},\
  \bibinfo {pages} {277206} (\bibinfo {year} {2021})}\BibitemShut {NoStop}%
\bibitem [{\citenamefont {Winter}\ \emph {et~al.}(2017)\citenamefont {Winter},
  \citenamefont {Tsirlin}, \citenamefont {Daghofer}, \citenamefont {Brink},
  \citenamefont {Singh}, \citenamefont {Gegenwart},\ and\ \citenamefont
  {Valentí}}]{winter17a}%
  \BibitemOpen
  \bibfield  {author} {\bibinfo {author} {\bibfnamefont {S.~M.}\ \bibnamefont
  {Winter}}, \bibinfo {author} {\bibfnamefont {A.~A.}\ \bibnamefont {Tsirlin}},
  \bibinfo {author} {\bibfnamefont {M.}~\bibnamefont {Daghofer}}, \bibinfo
  {author} {\bibfnamefont {J.~v.~d.}\ \bibnamefont {Brink}}, \bibinfo {author}
  {\bibfnamefont {Y.}~\bibnamefont {Singh}}, \bibinfo {author} {\bibfnamefont
  {P.}~\bibnamefont {Gegenwart}},\ and\ \bibinfo {author} {\bibfnamefont
  {R.}~\bibnamefont {Valentí}},\ }\bibfield  {title} {\bibinfo {title} {Models
  and materials for generalized {Kitaev} magnetism},\ }\href
  {https://doi.org/10.1088/1361-648X/aa8cf5} {\bibfield  {journal} {\bibinfo
  {journal} {Journal of Physics: Condensed Matter}\ }\textbf {\bibinfo {volume}
  {29}},\ \bibinfo {pages} {493002} (\bibinfo {year} {2017})}\BibitemShut
  {NoStop}%
\bibitem [{\citenamefont {Moessner}\ and\ \citenamefont
  {Chalker}(1998)}]{Moessner98_geo}%
  \BibitemOpen
  \bibfield  {author} {\bibinfo {author} {\bibfnamefont {R.}~\bibnamefont
  {Moessner}}\ and\ \bibinfo {author} {\bibfnamefont {J.~T.}\ \bibnamefont
  {Chalker}},\ }\bibfield  {title} {\bibinfo {title} {Low-temperature
  properties of classical geometrically frustrated antiferromagnets},\ }\href
  {https://doi.org/10.1103/PhysRevB.58.12049} {\bibfield  {journal} {\bibinfo
  {journal} {Physical Review B}\ }\textbf {\bibinfo {volume} {58}},\ \bibinfo
  {pages} {12049} (\bibinfo {year} {1998})}\BibitemShut {NoStop}%
\bibitem [{\citenamefont {Reimers}\ \emph {et~al.}(1991)\citenamefont
  {Reimers}, \citenamefont {Berlinsky},\ and\ \citenamefont {Shi}}]{Reimers91}%
  \BibitemOpen
  \bibfield  {author} {\bibinfo {author} {\bibfnamefont {J.~N.}\ \bibnamefont
  {Reimers}}, \bibinfo {author} {\bibfnamefont {A.~J.}\ \bibnamefont
  {Berlinsky}},\ and\ \bibinfo {author} {\bibfnamefont {A.-C.}\ \bibnamefont
  {Shi}},\ }\bibfield  {title} {\bibinfo {title} {Mean-field approach to
  magnetic ordering in highly frustrated pyrochlores},\ }\href
  {https://doi.org/10.1103/PhysRevB.43.865} {\bibfield  {journal} {\bibinfo
  {journal} {Physical Review B}\ }\textbf {\bibinfo {volume} {43}},\ \bibinfo
  {pages} {865} (\bibinfo {year} {1991})}\BibitemShut {NoStop}%
\bibitem [{\citenamefont {Nutakki}\ \emph {et~al.}(2023)\citenamefont
  {Nutakki}, \citenamefont {Jaubert},\ and\ \citenamefont
  {Pollet}}]{Nutakki2023}%
  \BibitemOpen
  \bibfield  {author} {\bibinfo {author} {\bibfnamefont {R.~P.}\ \bibnamefont
  {Nutakki}}, \bibinfo {author} {\bibfnamefont {L.~D.~C.}\ \bibnamefont
  {Jaubert}},\ and\ \bibinfo {author} {\bibfnamefont {L.}~\bibnamefont
  {Pollet}},\ }\bibfield  {title} {\bibinfo {title} {The {{Classical Heisenberg
  Model}} on the {{Centred Pyrochlore Lattice}}},\ }\bibfield  {journal}
  {\bibinfo  {journal} {arXiv:2303.11010 [cond-mat]}\ }\href
  {https://doi.org/10.48550/arXiv.2303.11010} {10.48550/arXiv.2303.11010}
  (\bibinfo {year} {2023})\BibitemShut {NoStop}%
\bibitem [{\citenamefont {Levin}(2002)}]{levin2002}%
  \BibitemOpen
  \bibfield  {author} {\bibinfo {author} {\bibfnamefont {Y.}~\bibnamefont
  {Levin}},\ }\bibfield  {title} {\bibinfo {title} {Electrostatic correlations:
  From plasma to biology},\ }\href
  {https://doi.org/10.1088/0034-4885/65/11/201} {\bibfield  {journal} {\bibinfo
   {journal} {Reports on Progress in Physics}\ }\textbf {\bibinfo {volume}
  {65}},\ \bibinfo {pages} {1577} (\bibinfo {year} {2002})}\BibitemShut
  {NoStop}%
\bibitem [{\citenamefont {Castelnovo}\ \emph {et~al.}(2011)\citenamefont
  {Castelnovo}, \citenamefont {Moessner},\ and\ \citenamefont
  {Sondhi}}]{castelnovo2011}%
  \BibitemOpen
  \bibfield  {author} {\bibinfo {author} {\bibfnamefont {C.}~\bibnamefont
  {Castelnovo}}, \bibinfo {author} {\bibfnamefont {R.}~\bibnamefont
  {Moessner}},\ and\ \bibinfo {author} {\bibfnamefont {S.~L.}\ \bibnamefont
  {Sondhi}},\ }\bibfield  {title} {\bibinfo {title} {Debye-{{H\"uckel}} theory
  for spin ice at low temperature},\ }\href
  {https://doi.org/10.1103/PhysRevB.84.144435} {\bibfield  {journal} {\bibinfo
  {journal} {Physical Review B}\ }\textbf {\bibinfo {volume} {84}},\ \bibinfo
  {pages} {144435} (\bibinfo {year} {2011})}\BibitemShut {NoStop}%
\bibitem [{\citenamefont {Kermarrec}\ \emph {et~al.}(2017)\citenamefont
  {Kermarrec}, \citenamefont {Gaudet}, \citenamefont {Fritsch}, \citenamefont
  {Khasanov}, \citenamefont {Guguchia}, \citenamefont {Ritter}, \citenamefont
  {Ross}, \citenamefont {Dabkowska},\ and\ \citenamefont
  {Gaulin}}]{Kermarrec2017}%
  \BibitemOpen
  \bibfield  {author} {\bibinfo {author} {\bibfnamefont {E.}~\bibnamefont
  {Kermarrec}}, \bibinfo {author} {\bibfnamefont {J.}~\bibnamefont {Gaudet}},
  \bibinfo {author} {\bibfnamefont {K.}~\bibnamefont {Fritsch}}, \bibinfo
  {author} {\bibfnamefont {R.}~\bibnamefont {Khasanov}}, \bibinfo {author}
  {\bibfnamefont {Z.}~\bibnamefont {Guguchia}}, \bibinfo {author}
  {\bibfnamefont {C.}~\bibnamefont {Ritter}}, \bibinfo {author} {\bibfnamefont
  {K.~A.}\ \bibnamefont {Ross}}, \bibinfo {author} {\bibfnamefont {H.~A.}\
  \bibnamefont {Dabkowska}},\ and\ \bibinfo {author} {\bibfnamefont {B.~D.}\
  \bibnamefont {Gaulin}},\ }\bibfield  {title} {\bibinfo {title} {Ground state
  selection under pressure in the quantum pyrochlore magnet yb2ti2o7},\ }\href
  {https://doi.org/10.1038/ncomms14810} {\bibfield  {journal} {\bibinfo
  {journal} {Nature Communications}\ }\textbf {\bibinfo {volume} {8}},\
  \bibinfo {pages} {14810} (\bibinfo {year} {2017})}\BibitemShut {NoStop}%
\bibitem [{\citenamefont {Zayed}\ \emph {et~al.}(2017)\citenamefont {Zayed},
  \citenamefont {R{\"u}egg}, \citenamefont {Larrea~J.}, \citenamefont
  {L{\"a}uchli}, \citenamefont {Panagopoulos}, \citenamefont {Saxena},
  \citenamefont {Ellerby}, \citenamefont {McMorrow}, \citenamefont
  {Str{\"a}ssle}, \citenamefont {Klotz}, \citenamefont {Hamel}, \citenamefont
  {Sadykov}, \citenamefont {Pomjakushin}, \citenamefont {Boehm}, \citenamefont
  {Jim{\'e}nez-Ruiz}, \citenamefont {Schneidewind}, \citenamefont
  {Pomjakushina}, \citenamefont {Stingaciu}, \citenamefont {Conder},\ and\
  \citenamefont {R{\o}nnow}}]{Zayed2017}%
  \BibitemOpen
  \bibfield  {author} {\bibinfo {author} {\bibfnamefont {M.~E.}\ \bibnamefont
  {Zayed}}, \bibinfo {author} {\bibfnamefont {C.}~\bibnamefont {R{\"u}egg}},
  \bibinfo {author} {\bibfnamefont {J.}~\bibnamefont {Larrea~J.}}, \bibinfo
  {author} {\bibfnamefont {A.~M.}\ \bibnamefont {L{\"a}uchli}}, \bibinfo
  {author} {\bibfnamefont {C.}~\bibnamefont {Panagopoulos}}, \bibinfo {author}
  {\bibfnamefont {S.~S.}\ \bibnamefont {Saxena}}, \bibinfo {author}
  {\bibfnamefont {M.}~\bibnamefont {Ellerby}}, \bibinfo {author} {\bibfnamefont
  {D.~F.}\ \bibnamefont {McMorrow}}, \bibinfo {author} {\bibfnamefont
  {T.}~\bibnamefont {Str{\"a}ssle}}, \bibinfo {author} {\bibfnamefont
  {S.}~\bibnamefont {Klotz}}, \bibinfo {author} {\bibfnamefont
  {G.}~\bibnamefont {Hamel}}, \bibinfo {author} {\bibfnamefont {R.~A.}\
  \bibnamefont {Sadykov}}, \bibinfo {author} {\bibfnamefont {V.}~\bibnamefont
  {Pomjakushin}}, \bibinfo {author} {\bibfnamefont {M.}~\bibnamefont {Boehm}},
  \bibinfo {author} {\bibfnamefont {M.}~\bibnamefont {Jim{\'e}nez-Ruiz}},
  \bibinfo {author} {\bibfnamefont {A.}~\bibnamefont {Schneidewind}}, \bibinfo
  {author} {\bibfnamefont {E.}~\bibnamefont {Pomjakushina}}, \bibinfo {author}
  {\bibfnamefont {M.}~\bibnamefont {Stingaciu}}, \bibinfo {author}
  {\bibfnamefont {K.}~\bibnamefont {Conder}},\ and\ \bibinfo {author}
  {\bibfnamefont {H.~M.}\ \bibnamefont {R{\o}nnow}},\ }\bibfield  {title}
  {\bibinfo {title} {4-spin plaquette singlet state in the shastry--sutherland
  compound srcu2(bo3)2},\ }\href {https://doi.org/10.1038/nphys4190} {\bibfield
   {journal} {\bibinfo  {journal} {Nature Physics}\ }\textbf {\bibinfo {volume}
  {13}},\ \bibinfo {pages} {962} (\bibinfo {year} {2017})}\BibitemShut
  {NoStop}%
\bibitem [{\citenamefont {Ross}\ \emph {et~al.}(2011)\citenamefont {Ross},
  \citenamefont {Savary}, \citenamefont {Gaulin},\ and\ \citenamefont
  {Balents}}]{Ross11}%
  \BibitemOpen
  \bibfield  {author} {\bibinfo {author} {\bibfnamefont {K.~A.}\ \bibnamefont
  {Ross}}, \bibinfo {author} {\bibfnamefont {L.}~\bibnamefont {Savary}},
  \bibinfo {author} {\bibfnamefont {B.~D.}\ \bibnamefont {Gaulin}},\ and\
  \bibinfo {author} {\bibfnamefont {L.}~\bibnamefont {Balents}},\ }\bibfield
  {title} {\bibinfo {title} {Quantum excitations in quantum spin ice},\ }\href
  {https://doi.org/10.1103/PhysRevX.1.021002} {\bibfield  {journal} {\bibinfo
  {journal} {Phys. Rev. X}\ }\textbf {\bibinfo {volume} {1}},\ \bibinfo {pages}
  {021002} (\bibinfo {year} {2011})}\BibitemShut {NoStop}%
\bibitem [{\citenamefont {R{\"o}{\ss}-Ohlenroth}\ \emph
  {et~al.}(2022)\citenamefont {R{\"o}{\ss}-Ohlenroth}, \citenamefont {Hirrle},
  \citenamefont {Kraft}, \citenamefont {Kalytta-Mewes}, \citenamefont {Jesche},
  \citenamefont {Krug~von Nidda},\ and\ \citenamefont
  {Volkmer}}]{RossOhlenroth2022}%
  \BibitemOpen
  \bibfield  {author} {\bibinfo {author} {\bibfnamefont {R.}~\bibnamefont
  {R{\"o}{\ss}-Ohlenroth}}, \bibinfo {author} {\bibfnamefont {M.}~\bibnamefont
  {Hirrle}}, \bibinfo {author} {\bibfnamefont {M.}~\bibnamefont {Kraft}},
  \bibinfo {author} {\bibfnamefont {A.}~\bibnamefont {Kalytta-Mewes}}, \bibinfo
  {author} {\bibfnamefont {A.}~\bibnamefont {Jesche}}, \bibinfo {author}
  {\bibfnamefont {H.-A.}\ \bibnamefont {Krug~von Nidda}},\ and\ \bibinfo
  {author} {\bibfnamefont {D.}~\bibnamefont {Volkmer}},\ }\bibfield  {title}
  {\bibinfo {title} {Synthesis, thermal stability and magnetic properties of an
  interpenetrated mn(ii) triazolate coordination framework},\ }\href
  {https://doi.org/https://doi.org/10.1002/zaac.202200153} {\bibfield
  {journal} {\bibinfo  {journal} {Zeitschrift f{\"u}r anorganische und
  allgemeine Chemie}\ }\textbf {\bibinfo {volume} {648}},\ \bibinfo {pages}
  {e202200153} (\bibinfo {year} {2022})}\BibitemShut {NoStop}%
\bibitem [{\citenamefont {Gaenko}\ \emph {et~al.}(2017)\citenamefont {Gaenko},
  \citenamefont {Antipov}, \citenamefont {Carcassi}, \citenamefont {Chen},
  \citenamefont {Chen}, \citenamefont {Dong}, \citenamefont {Gamper},
  \citenamefont {Gukelberger}, \citenamefont {Igarashi}, \citenamefont
  {Iskakov}, \citenamefont {K{\"o}nz}, \citenamefont {LeBlanc}, \citenamefont
  {Levy}, \citenamefont {Ma}, \citenamefont {Paki}, \citenamefont {Shinaoka},
  \citenamefont {Todo}, \citenamefont {Troyer},\ and\ \citenamefont
  {Gull}}]{ALPSCore}%
  \BibitemOpen
  \bibfield  {author} {\bibinfo {author} {\bibfnamefont {A.}~\bibnamefont
  {Gaenko}}, \bibinfo {author} {\bibfnamefont {A.}~\bibnamefont {Antipov}},
  \bibinfo {author} {\bibfnamefont {G.}~\bibnamefont {Carcassi}}, \bibinfo
  {author} {\bibfnamefont {T.}~\bibnamefont {Chen}}, \bibinfo {author}
  {\bibfnamefont {X.}~\bibnamefont {Chen}}, \bibinfo {author} {\bibfnamefont
  {Q.}~\bibnamefont {Dong}}, \bibinfo {author} {\bibfnamefont {L.}~\bibnamefont
  {Gamper}}, \bibinfo {author} {\bibfnamefont {J.}~\bibnamefont {Gukelberger}},
  \bibinfo {author} {\bibfnamefont {R.}~\bibnamefont {Igarashi}}, \bibinfo
  {author} {\bibfnamefont {S.}~\bibnamefont {Iskakov}}, \bibinfo {author}
  {\bibfnamefont {M.}~\bibnamefont {K{\"o}nz}}, \bibinfo {author}
  {\bibfnamefont {J.}~\bibnamefont {LeBlanc}}, \bibinfo {author} {\bibfnamefont
  {R.}~\bibnamefont {Levy}}, \bibinfo {author} {\bibfnamefont {P.}~\bibnamefont
  {Ma}}, \bibinfo {author} {\bibfnamefont {J.}~\bibnamefont {Paki}}, \bibinfo
  {author} {\bibfnamefont {H.}~\bibnamefont {Shinaoka}}, \bibinfo {author}
  {\bibfnamefont {S.}~\bibnamefont {Todo}}, \bibinfo {author} {\bibfnamefont
  {M.}~\bibnamefont {Troyer}},\ and\ \bibinfo {author} {\bibfnamefont
  {E.}~\bibnamefont {Gull}},\ }\bibfield  {title} {\bibinfo {title} {{Updated
  core libraries of the ALPS project}},\ }\href
  {https://doi.org/10.1016/j.cpc.2016.12.009} {\bibfield  {journal} {\bibinfo
  {journal} {Comput. Phys. Commun.}\ }\textbf {\bibinfo {volume} {213}},\
  \bibinfo {pages} {235} (\bibinfo {year} {2017})}\BibitemShut {NoStop}%
\end{thebibliography}%


\begin{thebibliography}{23}%
\makeatletter
\providecommand \@ifxundefined [1]{%
 \@ifx{#1\undefined}
}%
\providecommand \@ifnum [1]{%
 \ifnum #1\expandafter \@firstoftwo
 \else \expandafter \@secondoftwo
 \fi
}%
\providecommand \@ifx [1]{%
 \ifx #1\expandafter \@firstoftwo
 \else \expandafter \@secondoftwo
 \fi
}%
\providecommand \natexlab [1]{#1}%
\providecommand \enquote  [1]{``#1''}%
\providecommand \bibnamefont  [1]{#1}%
\providecommand \bibfnamefont [1]{#1}%
\providecommand \citenamefont [1]{#1}%
\providecommand \href@noop [0]{\@secondoftwo}%
\providecommand \href [0]{\begingroup \@sanitize@url \@href}%
\providecommand \@href[1]{\@@startlink{#1}\@@href}%
\providecommand \@@href[1]{\endgroup#1\@@endlink}%
\providecommand \@sanitize@url [0]{\catcode `\\12\catcode `\$12\catcode
  `\&12\catcode `\#12\catcode `\^12\catcode `\_12\catcode `\%12\relax}%
\providecommand \@@startlink[1]{}%
\providecommand \@@endlink[0]{}%
\providecommand \url  [0]{\begingroup\@sanitize@url \@url }%
\providecommand \@url [1]{\endgroup\@href {#1}{\urlprefix }}%
\providecommand \urlprefix  [0]{URL }%
\providecommand \Eprint [0]{\href }%
\providecommand \doibase [0]{https://doi.org/}%
\providecommand \selectlanguage [0]{\@gobble}%
\providecommand \bibinfo  [0]{\@secondoftwo}%
\providecommand \bibfield  [0]{\@secondoftwo}%
\providecommand \translation [1]{[#1]}%
\providecommand \BibitemOpen [0]{}%
\providecommand \bibitemStop [0]{}%
\providecommand \bibitemNoStop [0]{.\EOS\space}%
\providecommand \EOS [0]{\spacefactor3000\relax}%
\providecommand \BibitemShut  [1]{\csname bibitem#1\endcsname}%
\let\auto@bib@innerbib\@empty
\bibitem [{\citenamefont {Gándara}\ \emph {et~al.}(2012)\citenamefont
  {Gándara}, \citenamefont {Uribe-Romo}, \citenamefont {Britt}, \citenamefont
  {Furukawa}, \citenamefont {Lei}, \citenamefont {Cheng}, \citenamefont {Duan},
  \citenamefont {O'Keeffe},\ and\ \citenamefont {Yaghi}}]{gandara12}%
  \BibitemOpen
  \bibfield  {author} {\bibinfo {author} {\bibfnamefont {F.}~\bibnamefont
  {Gándara}}, \bibinfo {author} {\bibfnamefont {F.~J.}\ \bibnamefont
  {Uribe-Romo}}, \bibinfo {author} {\bibfnamefont {D.~K.}\ \bibnamefont
  {Britt}}, \bibinfo {author} {\bibfnamefont {H.}~\bibnamefont {Furukawa}},
  \bibinfo {author} {\bibfnamefont {L.}~\bibnamefont {Lei}}, \bibinfo {author}
  {\bibfnamefont {R.}~\bibnamefont {Cheng}}, \bibinfo {author} {\bibfnamefont
  {X.}~\bibnamefont {Duan}}, \bibinfo {author} {\bibfnamefont {M.}~\bibnamefont
  {O'Keeffe}},\ and\ \bibinfo {author} {\bibfnamefont {O.~M.}\ \bibnamefont
  {Yaghi}},\ }\href {https://doi.org/10.1002/chem.201103433} {\bibfield
  {journal} {\bibinfo  {journal} {Chemistry - A European Journal}\ }\textbf
  {\bibinfo {volume} {18}},\ \bibinfo {pages} {10595} (\bibinfo {year}
  {2012})}\BibitemShut {NoStop}%
\bibitem [{\citenamefont {Pet{\u r}{\'\i}{\u c}ek}\ \emph
  {et~al.}(2014)\citenamefont {Pet{\u r}{\'\i}{\u c}ek}, \citenamefont {Du{\u
  s}ek},\ and\ \citenamefont {Palatinus}}]{jana2006}%
  \BibitemOpen
  \bibfield  {author} {\bibinfo {author} {\bibfnamefont {V.}~\bibnamefont
  {Pet{\u r}{\'\i}{\u c}ek}}, \bibinfo {author} {\bibfnamefont
  {M.}~\bibnamefont {Du{\u s}ek}},\ and\ \bibinfo {author} {\bibfnamefont
  {L.}~\bibnamefont {Palatinus}},\ }\href
  {https://doi.org/10.1515/zkri-2014-1737} {\bibfield  {journal} {\bibinfo
  {journal} {Z. Krist.}\ }\textbf {\bibinfo {volume} {229}},\ \bibinfo {pages}
  {345} (\bibinfo {year} {2014})}\BibitemShut {NoStop}%
\bibitem [{\citenamefont {Koepernik}\ and\ \citenamefont
  {Eschrig}(1999)}]{Kopernik99}%
  \BibitemOpen
  \bibfield  {author} {\bibinfo {author} {\bibfnamefont {K.}~\bibnamefont
  {Koepernik}}\ and\ \bibinfo {author} {\bibfnamefont {H.}~\bibnamefont
  {Eschrig}},\ }\href {https://doi.org/10.1103/PhysRevB.59.1743} {\bibfield
  {journal} {\bibinfo  {journal} {Phys. Rev. B}\ }\textbf {\bibinfo {volume}
  {59}},\ \bibinfo {pages} {1743} (\bibinfo {year} {1999})}\BibitemShut
  {NoStop}%
\bibitem [{\citenamefont {He}\ \emph {et~al.}(2017)\citenamefont {He},
  \citenamefont {Ye}, \citenamefont {Xu}, \citenamefont {Zhou}, \citenamefont
  {Zhou}, \citenamefont {Chen}, \citenamefont {Zhang},\ and\ \citenamefont
  {Chen}}]{He17}%
  \BibitemOpen
  \bibfield  {author} {\bibinfo {author} {\bibfnamefont {C.-T.}\ \bibnamefont
  {He}}, \bibinfo {author} {\bibfnamefont {Z.-M.}\ \bibnamefont {Ye}}, \bibinfo
  {author} {\bibfnamefont {Y.-T.}\ \bibnamefont {Xu}}, \bibinfo {author}
  {\bibfnamefont {D.-D.}\ \bibnamefont {Zhou}}, \bibinfo {author}
  {\bibfnamefont {H.-L.}\ \bibnamefont {Zhou}}, \bibinfo {author}
  {\bibfnamefont {D.}~\bibnamefont {Chen}}, \bibinfo {author} {\bibfnamefont
  {J.-P.}\ \bibnamefont {Zhang}},\ and\ \bibinfo {author} {\bibfnamefont
  {X.-M.}\ \bibnamefont {Chen}},\ }\href {https://doi.org/10.1039/C7SC03067C}
  {\bibfield  {journal} {\bibinfo  {journal} {Chem. Sci.}\ }\textbf {\bibinfo
  {volume} {8}},\ \bibinfo {pages} {7560} (\bibinfo {year} {2017})}\BibitemShut
  {NoStop}%
\bibitem [{\citenamefont {Xiang}\ \emph {et~al.}(2011)\citenamefont {Xiang},
  \citenamefont {Kan}, \citenamefont {Wei}, \citenamefont {Whangbo},\ and\
  \citenamefont {Gong}}]{Xiang11}%
  \BibitemOpen
  \bibfield  {author} {\bibinfo {author} {\bibfnamefont {H.~J.}\ \bibnamefont
  {Xiang}}, \bibinfo {author} {\bibfnamefont {E.~J.}\ \bibnamefont {Kan}},
  \bibinfo {author} {\bibfnamefont {S.-H.}\ \bibnamefont {Wei}}, \bibinfo
  {author} {\bibfnamefont {M.-H.}\ \bibnamefont {Whangbo}},\ and\ \bibinfo
  {author} {\bibfnamefont {X.~G.}\ \bibnamefont {Gong}},\ }\href
  {https://doi.org/10.1103/PhysRevB.84.224429} {\bibfield  {journal} {\bibinfo
  {journal} {Phys. Rev. B}\ }\textbf {\bibinfo {volume} {84}},\ \bibinfo
  {pages} {224429} (\bibinfo {year} {2011})}\BibitemShut {NoStop}%
\bibitem [{\citenamefont {Perdew}\ \emph {et~al.}(1996)\citenamefont {Perdew},
  \citenamefont {Burke},\ and\ \citenamefont {Ernzerhof}}]{Perdew96}%
  \BibitemOpen
  \bibfield  {author} {\bibinfo {author} {\bibfnamefont {J.~P.}\ \bibnamefont
  {Perdew}}, \bibinfo {author} {\bibfnamefont {K.}~\bibnamefont {Burke}},\ and\
  \bibinfo {author} {\bibfnamefont {M.}~\bibnamefont {Ernzerhof}},\ }\href
  {https://doi.org/10.1103/PhysRevLett.77.3865} {\bibfield  {journal} {\bibinfo
   {journal} {Phys. Rev. Lett.}\ }\textbf {\bibinfo {volume} {77}},\ \bibinfo
  {pages} {3865} (\bibinfo {year} {1996})}\BibitemShut {NoStop}%
\bibitem [{\citenamefont {Nath}\ \emph {et~al.}(2014)\citenamefont {Nath},
  \citenamefont {Ranjith}, \citenamefont {Roy}, \citenamefont {Johnston},
  \citenamefont {Furukawa},\ and\ \citenamefont {Tsirlin}}]{Nath14}%
  \BibitemOpen
  \bibfield  {author} {\bibinfo {author} {\bibfnamefont {R.}~\bibnamefont
  {Nath}}, \bibinfo {author} {\bibfnamefont {K.~M.}\ \bibnamefont {Ranjith}},
  \bibinfo {author} {\bibfnamefont {B.}~\bibnamefont {Roy}}, \bibinfo {author}
  {\bibfnamefont {D.~C.}\ \bibnamefont {Johnston}}, \bibinfo {author}
  {\bibfnamefont {Y.}~\bibnamefont {Furukawa}},\ and\ \bibinfo {author}
  {\bibfnamefont {A.~A.}\ \bibnamefont {Tsirlin}},\ }\href
  {https://doi.org/10.1103/PhysRevB.90.024431} {\bibfield  {journal} {\bibinfo
  {journal} {Phys. Rev. B}\ }\textbf {\bibinfo {volume} {90}},\ \bibinfo
  {pages} {024431} (\bibinfo {year} {2014})}\BibitemShut {NoStop}%
\bibitem [{\citenamefont {Katsura}\ \emph {et~al.}(2010)\citenamefont
  {Katsura}, \citenamefont {Maruyama}, \citenamefont {Tanaka},\ and\
  \citenamefont {Tasaki}}]{katsura10}%
  \BibitemOpen
  \bibfield  {author} {\bibinfo {author} {\bibfnamefont {H.}~\bibnamefont
  {Katsura}}, \bibinfo {author} {\bibfnamefont {I.}~\bibnamefont {Maruyama}},
  \bibinfo {author} {\bibfnamefont {A.}~\bibnamefont {Tanaka}},\ and\ \bibinfo
  {author} {\bibfnamefont {H.}~\bibnamefont {Tasaki}},\ }\href
  {https://doi.org/10.1209/0295-5075/91/57007} {\bibfield  {journal} {\bibinfo
  {journal} {Europhysics Letters}\ }\textbf {\bibinfo {volume} {91}},\ \bibinfo
  {pages} {57007} (\bibinfo {year} {2010})}\BibitemShut {NoStop}%
\bibitem [{\citenamefont {Essafi}\ \emph {et~al.}(2017)\citenamefont {Essafi},
  \citenamefont {Jaubert},\ and\ \citenamefont {Udagawa}}]{essafi17}%
  \BibitemOpen
  \bibfield  {author} {\bibinfo {author} {\bibfnamefont {K.}~\bibnamefont
  {Essafi}}, \bibinfo {author} {\bibfnamefont {L.~D.~C.}\ \bibnamefont
  {Jaubert}},\ and\ \bibinfo {author} {\bibfnamefont {M.}~\bibnamefont
  {Udagawa}},\ }\href {https://doi.org/10.1088/1361-648X/aa782f} {\bibfield
  {journal} {\bibinfo  {journal} {Journal of Physics: Condensed Matter}\
  }\textbf {\bibinfo {volume} {29}},\ \bibinfo {pages} {315802} (\bibinfo
  {year} {2017})}\BibitemShut {NoStop}%
\bibitem [{\citenamefont {Mathai}\ and\ \citenamefont
  {Haubold}(2017)}]{mathai_2017}%
  \BibitemOpen
  \bibfield  {author} {\bibinfo {author} {\bibfnamefont {A.~M.}\ \bibnamefont
  {Mathai}}\ and\ \bibinfo {author} {\bibfnamefont {H.~J.}\ \bibnamefont
  {Haubold}},\ }\href@noop {} {\emph {\bibinfo {title} {Linear {Algebra}}}},\
  De {Gruyter} {Textbook}\ (\bibinfo  {publisher} {Walter de Gruyter},\
  \bibinfo {year} {2017})\BibitemShut {NoStop}%
\bibitem [{\citenamefont {Katznelson}\ and\ \citenamefont
  {Katznelson}(2007)}]{katznelson_2007}%
  \BibitemOpen
  \bibfield  {author} {\bibinfo {author} {\bibfnamefont {Y.}~\bibnamefont
  {Katznelson}}\ and\ \bibinfo {author} {\bibfnamefont {Y.}~\bibnamefont
  {Katznelson}},\ }\href {https://doi.org/10.1090/stml/044} {\emph {\bibinfo
  {title} {A ({Terse}) {Introduction} to {Linear} {Algebra}}}},\ \bibinfo
  {series} {The {Student} {Mathematical} {Library}}, Vol.~\bibinfo {volume}
  {44}\ (\bibinfo  {publisher} {American Mathematical Society},\ \bibinfo
  {address} {Providence, Rhode Island},\ \bibinfo {year} {2007})\BibitemShut
  {NoStop}%
\bibitem [{\citenamefont {Lyons}\ and\ \citenamefont {Kaplan}(1960)}]{lyons60}%
  \BibitemOpen
  \bibfield  {author} {\bibinfo {author} {\bibfnamefont {D.~H.}\ \bibnamefont
  {Lyons}}\ and\ \bibinfo {author} {\bibfnamefont {T.~A.}\ \bibnamefont
  {Kaplan}},\ }\href {https://doi.org/10.1103/PhysRev.120.1580} {\bibfield
  {journal} {\bibinfo  {journal} {Physical Review}\ }\textbf {\bibinfo {volume}
  {120}},\ \bibinfo {pages} {1580} (\bibinfo {year} {1960})}\BibitemShut
  {NoStop}%
\bibitem [{\citenamefont {Nutakki}\ \emph {et~al.}(2023)\citenamefont
  {Nutakki}, \citenamefont {Jaubert},\ and\ \citenamefont
  {Pollet}}]{Nutakki2023}%
  \BibitemOpen
  \bibfield  {author} {\bibinfo {author} {\bibfnamefont {R.~P.}\ \bibnamefont
  {Nutakki}}, \bibinfo {author} {\bibfnamefont {L.~D.~C.}\ \bibnamefont
  {Jaubert}},\ and\ \bibinfo {author} {\bibfnamefont {L.}~\bibnamefont
  {Pollet}},\ }\bibfield  {journal} {\bibinfo  {journal} {arXiv:2303.11010
  [cond-mat]}\ }\href {https://doi.org/10.48550/arXiv.2303.11010}
  {10.48550/arXiv.2303.11010} (\bibinfo {year} {2023})\BibitemShut {NoStop}%
\bibitem [{\citenamefont {Creutz}(1980)}]{creutz80}%
  \BibitemOpen
  \bibfield  {author} {\bibinfo {author} {\bibfnamefont {M.}~\bibnamefont
  {Creutz}},\ }\href {https://doi.org/10.1103/PhysRevD.21.2308} {\bibfield
  {journal} {\bibinfo  {journal} {Physical Review D}\ }\textbf {\bibinfo
  {volume} {21}},\ \bibinfo {pages} {2308} (\bibinfo {year}
  {1980})}\BibitemShut {NoStop}%
\bibitem [{\citenamefont {Janke}(2008)}]{janke08}%
  \BibitemOpen
  \bibfield  {author} {\bibinfo {author} {\bibfnamefont {W.}~\bibnamefont
  {Janke}},\ }in\ \href {https://doi.org/10.1007/978-3-540-74686-7_4} {\emph
  {\bibinfo {booktitle} {Computational {Many}-{Particle} {Physics}}}},\ Vol.\
  \bibinfo {volume} {739},\ \bibinfo {editor} {edited by\ \bibinfo {editor}
  {\bibfnamefont {H.}~\bibnamefont {Fehske}}, \bibinfo {editor} {\bibfnamefont
  {R.}~\bibnamefont {Schneider}},\ and\ \bibinfo {editor} {\bibfnamefont
  {A.}~\bibnamefont {Weiße}}}\ (\bibinfo  {publisher} {Springer Berlin
  Heidelberg},\ \bibinfo {address} {Berlin, Heidelberg},\ \bibinfo {year}
  {2008})\ pp.\ \bibinfo {pages} {79--140}\BibitemShut {NoStop}%
\bibitem [{\citenamefont {Greitemann}(2019)}]{greitemann19}%
  \BibitemOpen
  \bibfield  {author} {\bibinfo {author} {\bibfnamefont {J.~F.}\ \bibnamefont
  {Greitemann}},\ }\emph {\bibinfo {title} {Investigation of {Hidden}
  {Multipolar} {Spin} {Order} in {Frustrated} {Magnets} {Using} {Interpretable}
  {Machine} {Learning} {Techniqes}}},\ \href@noop {} {Ph.D. thesis},\ \bibinfo
  {school} {LMU Munich} (\bibinfo {year} {2019})\BibitemShut {NoStop}%
\bibitem [{\citenamefont {Brown}\ and\ \citenamefont {Woch}(1987)}]{brown87}%
  \BibitemOpen
  \bibfield  {author} {\bibinfo {author} {\bibfnamefont {F.~R.}\ \bibnamefont
  {Brown}}\ and\ \bibinfo {author} {\bibfnamefont {T.~J.}\ \bibnamefont
  {Woch}},\ }\href {https://doi.org/10.1103/PhysRevLett.58.2394} {\bibfield
  {journal} {\bibinfo  {journal} {Physical Review Letters}\ }\textbf {\bibinfo
  {volume} {58}},\ \bibinfo {pages} {2394} (\bibinfo {year}
  {1987})}\BibitemShut {NoStop}%
\bibitem [{\citenamefont {Creutz}(1987)}]{creutz87}%
  \BibitemOpen
  \bibfield  {author} {\bibinfo {author} {\bibfnamefont {M.}~\bibnamefont
  {Creutz}},\ }\href {https://doi.org/10.1103/PhysRevD.36.515} {\bibfield
  {journal} {\bibinfo  {journal} {Physical Review D}\ }\textbf {\bibinfo
  {volume} {36}},\ \bibinfo {pages} {515} (\bibinfo {year} {1987})}\BibitemShut
  {NoStop}%
\bibitem [{\citenamefont {Landau}\ and\ \citenamefont
  {Binder}(2009)}]{landau09}%
  \BibitemOpen
  \bibfield  {author} {\bibinfo {author} {\bibfnamefont {D.~P.}\ \bibnamefont
  {Landau}}\ and\ \bibinfo {author} {\bibfnamefont {K.}~\bibnamefont
  {Binder}},\ }\href@noop {} {\emph {\bibinfo {title} {A {Guide} to {Monte}
  {Carlo} {Simulations} in {Statistical} {Physics}, {Third} {Edition}}}},\
  \bibinfo {edition} {3rd}\ ed.\ (\bibinfo  {publisher} {Cambridge University
  Press},\ \bibinfo {address} {Cambridge},\ \bibinfo {year} {2009})\BibitemShut
  {NoStop}%
\bibitem [{\citenamefont {Isakov}\ \emph {et~al.}(2004)\citenamefont {Isakov},
  \citenamefont {Gregor}, \citenamefont {Moessner},\ and\ \citenamefont
  {Sondhi}}]{isakov04}%
  \BibitemOpen
  \bibfield  {author} {\bibinfo {author} {\bibfnamefont {S.~V.}\ \bibnamefont
  {Isakov}}, \bibinfo {author} {\bibfnamefont {K.}~\bibnamefont {Gregor}},
  \bibinfo {author} {\bibfnamefont {R.}~\bibnamefont {Moessner}},\ and\
  \bibinfo {author} {\bibfnamefont {S.~L.}\ \bibnamefont {Sondhi}},\ }\href
  {https://doi.org/10.1103/PhysRevLett.93.167204} {\bibfield  {journal}
  {\bibinfo  {journal} {Physical Review Letters}\ }\textbf {\bibinfo {volume}
  {93}},\ \bibinfo {pages} {167204} (\bibinfo {year} {2004})}\BibitemShut
  {NoStop}%
\bibitem [{\citenamefont {Henley}(2005)}]{henley05}%
  \BibitemOpen
  \bibfield  {author} {\bibinfo {author} {\bibfnamefont {C.~L.}\ \bibnamefont
  {Henley}},\ }\href {https://doi.org/10.1103/PhysRevB.71.014424} {\bibfield
  {journal} {\bibinfo  {journal} {Physical Review B}\ }\textbf {\bibinfo
  {volume} {71}},\ \bibinfo {pages} {014424} (\bibinfo {year}
  {2005})}\BibitemShut {NoStop}%
\bibitem [{\citenamefont {Henley}(2010)}]{henley10}%
  \BibitemOpen
  \bibfield  {author} {\bibinfo {author} {\bibfnamefont {C.~L.}\ \bibnamefont
  {Henley}},\ }\href {https://doi.org/10.1146/annurev-conmatphys-070909-104138}
  {\bibfield  {journal} {\bibinfo  {journal} {Annu. Rev. Condens. Matter
  Phys.}\ }\textbf {\bibinfo {volume} {1}},\ \bibinfo {pages} {179} (\bibinfo
  {year} {2010})}\BibitemShut {NoStop}%
\bibitem [{\citenamefont {Yan}\ \emph {et~al.}(2017)\citenamefont {Yan},
  \citenamefont {Benton}, \citenamefont {Jaubert},\ and\ \citenamefont
  {Shannon}}]{Yan2017}%
  \BibitemOpen
  \bibfield  {author} {\bibinfo {author} {\bibfnamefont {H.}~\bibnamefont
  {Yan}}, \bibinfo {author} {\bibfnamefont {O.}~\bibnamefont {Benton}},
  \bibinfo {author} {\bibfnamefont {L.}~\bibnamefont {Jaubert}},\ and\ \bibinfo
  {author} {\bibfnamefont {N.}~\bibnamefont {Shannon}},\ }\href
  {https://doi.org/10.1103/PhysRevB.95.094422} {\bibfield  {journal} {\bibinfo
  {journal} {Physical Review B}\ }\textbf {\bibinfo {volume} {95}},\ \bibinfo
  {pages} {094422} (\bibinfo {year} {2017})}\BibitemShut {NoStop}%
\end{thebibliography}%
\end{document}


\title{Supplementary Information for: ``Frustration on a centred pyrochlore lattice in metal-organic frameworks"}
	\author{Rajah Nutakki}
	\affiliation{Arnold Sommerfeld Center for Theoretical Physics, University of Munich, Theresienstr. 37, 80333 M\"unchen, Germany}
	\affiliation{Munich Center for Quantum Science and Technology (MCQST), Schellingstr. 4, 80799 M\"unchen, Germany}
	
	\author{Richard R\"o\ss-Ohlenroth}
	\affiliation{Chair of Solid State and Materials Chemistry, Institute of Physics, University of Augsburg, D-86159 Augsburg, Germany}
	
	\author{Dirk Volkmer}
	\affiliation{Chair of Solid State and Materials Chemistry, Institute of Physics, University of Augsburg, D-86159 Augsburg, Germany}
	
	\author{Anton Jesche}
	\affiliation{Experimental Physics VI, Center for Electronic Correlations and Magnetism, Institute of Physics, University of Augsburg, 86159 Augsburg, Germany}
	
	\author{Hans-Albrecht Krug von Nidda}
	\affiliation{Experimental Physics V, Center for Electronic Correlations and Magnetism, Institute of Physics, University of Augsburg, 86159 Augsburg, Germany}
	
	\author{Alexander A. Tsirlin}
	\affiliation{Experimental Physics VI, Center for Electronic Correlations and Magnetism, Institute of Physics, University of Augsburg, 86159 Augsburg, Germany}
	
	\author{Philipp Gegenwart}
	\affiliation{Experimental Physics VI, Center for Electronic Correlations and Magnetism, Institute of Physics, University of Augsburg, 86159 Augsburg, Germany}
	
	\author{Lode Pollet}
	\affiliation{Arnold Sommerfeld Center for Theoretical Physics, University of Munich, Theresienstr. 37, 80333 M\"unchen, Germany}
	\affiliation{Munich Center for Quantum Science and Technology (MCQST), Schellingstr. 4, 80799 M\"unchen, Germany}
	
	\author{Ludovic D.\ C.\ Jaubert}
	\affiliation{CNRS, Universit\'e de Bordeaux, LOMA, UMR 5798, 33400 Talence, France}
	\date{\today}
\maketitle
\section{Synthesis and Crystallographic Characterisation}
[Mn(ta)$_{2}$] and [Zn(ta)$_{2}$] were prepared by procedures adapted from literature \cite{gandara12} and characterised by IR-spectroscopy, x-ray powder diffraction (XRPD) and electron microscopy. The synthesis of [Mn(ta)$_{2}$] was conducted twice in order to re-evaluate the magnetic properties and rule out batch effects.
\subsection{Materials}
$N,N$-Diethylformamide (99\%; TCI), $N,N$-dimethylformamide (99.8\% analytical grade; VWR), ethanol (99.8\% analytical grade; VWR), methanol (99.8\% analytical grade; VWR), ammonia solution (25\% technical grade, VWR), 1$H$-1,2,3-triazole ($H$-ta; 98\%; BLD Pharmatech Ltd.), zinc(II) chloride (99.999\%; Sigma-Aldrich) and manganese(II) nitrate hydrate (99.995\%; Alfa Aesar) were used as received from the commercial supplier.
\subsection{Pellet Preparation}
Pellets for the measurements of both compounds were prepared with a 3 mm diameter Maassen press die set, applying a force of ca. 2 kN for 30 minutes. The pellets were dried again for 1-2 h under vacuum and kept under Argon atmosphere prior to measurements. The structural integrity after pressing of both compounds was verified with the XRPD patterns of ground pellets.
\subsection{XRPD measurements}
\begin{figure}[h]
	\centering
	\includegraphics[width=7.5cm]{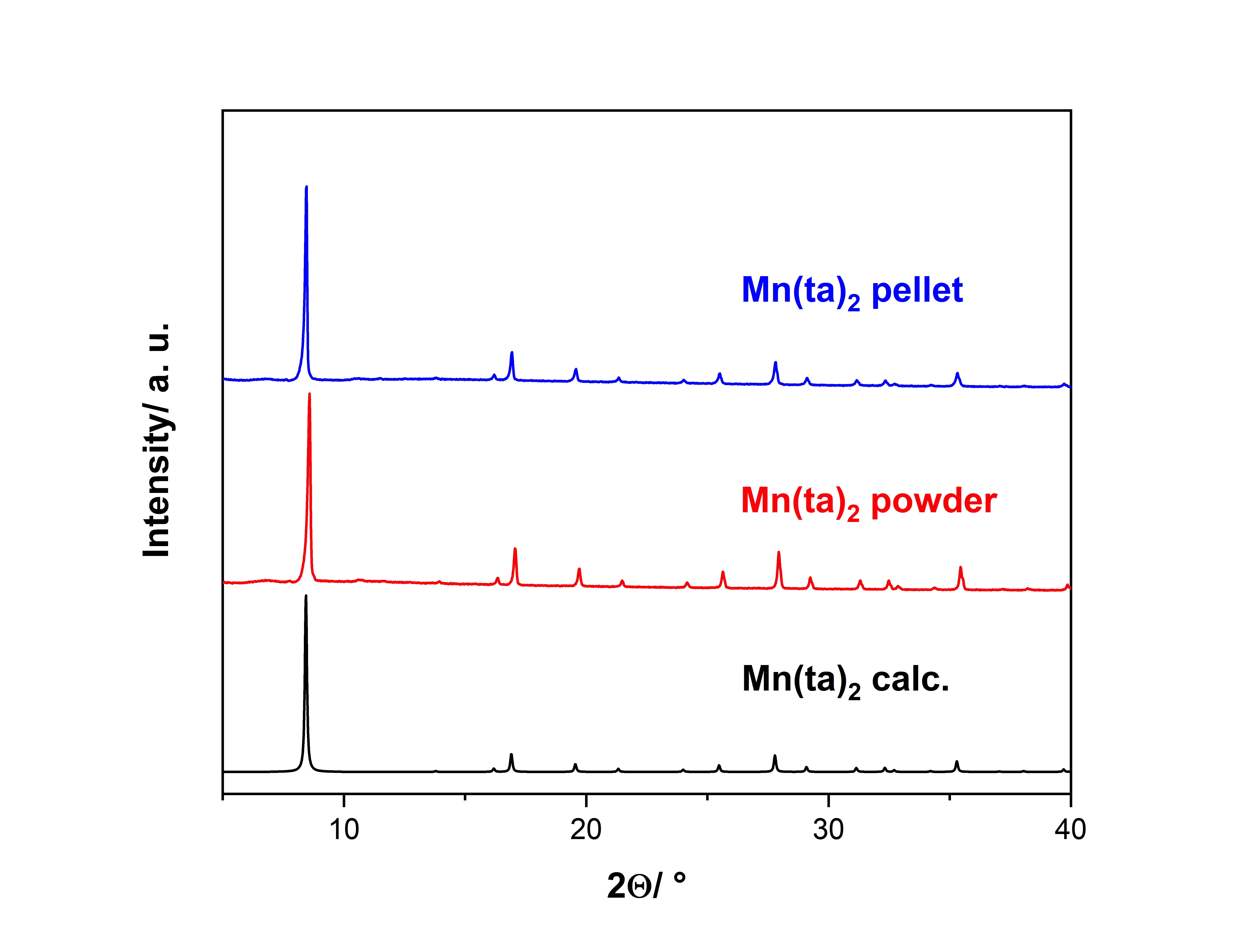}
	\includegraphics[width=7.5cm]{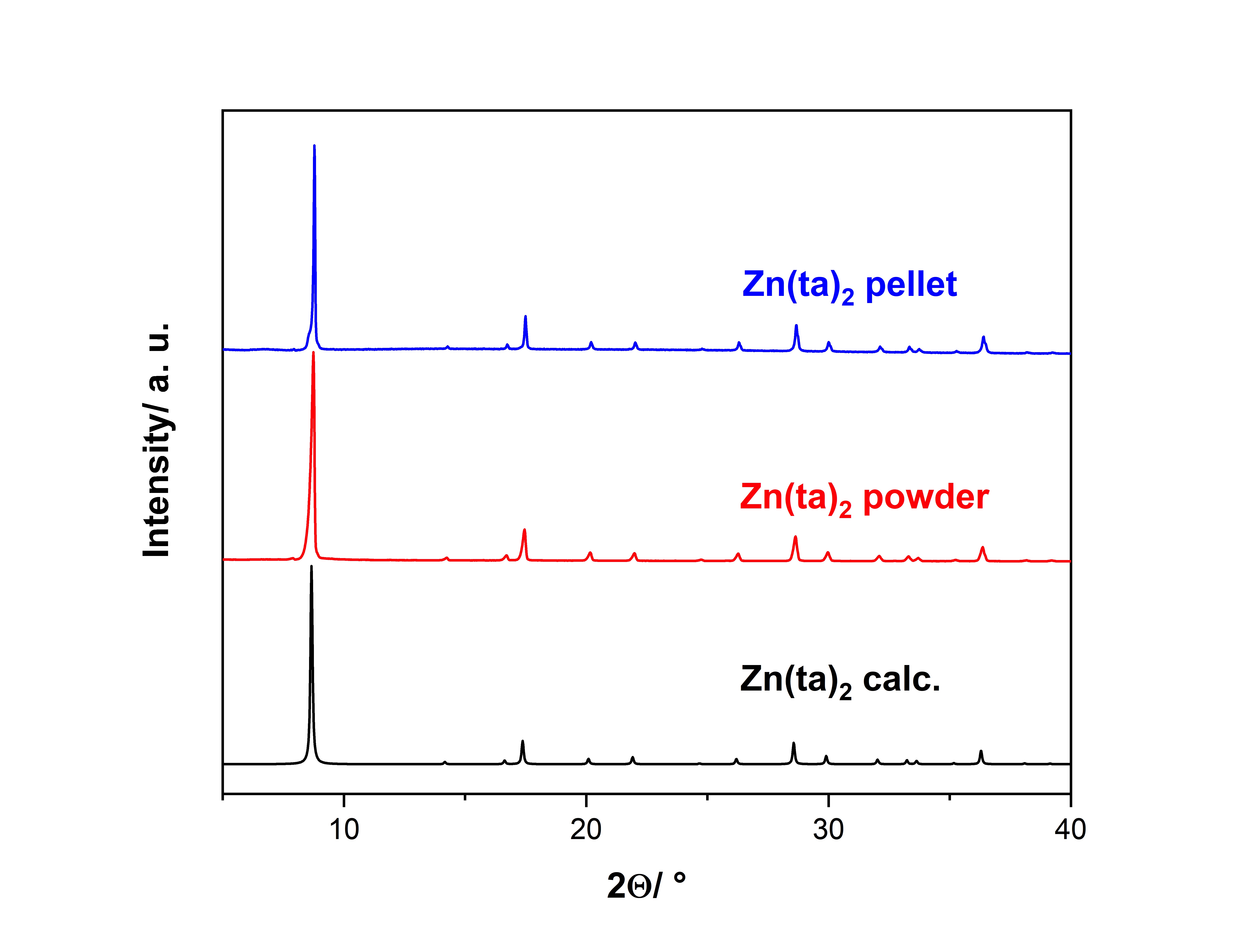}
	\caption{
		\textit{Left:} Room-temperature x-ray powder diffraction (XRPD) pattern for Mn(ta)$_{2}$ calculated from literature (black) \cite{gandara12}, the powder sample (red) and the pellet sample (blue).
		\textit{Right:} Room-temperature x-ray powder diffraction (XRPD) pattern for Zn(ta)$_{2}$ calculated from literature (black) \cite{gandara12}, the powder sample (red) and the pellet sample (blue).
	}
	\label{fig:XRPD}
\end{figure}

\begin{figure}
	\includegraphics[width=9cm]{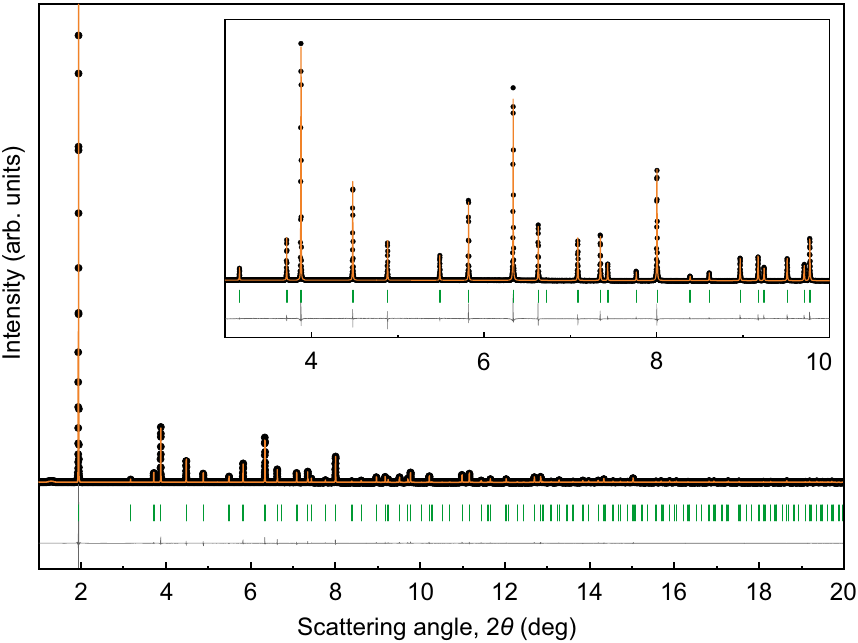}
	\caption{Rietveld refinement for the high-resolution XRD data collected at 5\,K for Mn(ta)$_2$. 
	}
	\label{fig:XRPD-5K}
\end{figure}

A Seifert XRD 3003 TT powder diffractometer with a Meteor1D detector was used to measure the XRPD patterns at room temperature in the 4-40$^\circ$ 2$\theta$ range using Cu K$\alpha_{1}$ radiation, see fig. \ref{fig:XRPD}.
Additionally, we performed the high-resolution XRPD measurement at 5\,K to confirm the cubic symmetry of Mn(ta)$_{2}$ at low temperatures, see fig. \ref{fig:XRPD-5K}. The data were collected at the ID22 beamline of the European Synchrotron Radiation Facility (ESRF, Grenoble) using the wavelength of 0.35432\,\r A and the multi-analyzer detector setup. The powder sample was loaded into the glass capillary and spun during the measurement. The \texttt{Jana2006} software was used for the Rietveld refinement~\cite{jana2006}. The refined structural parameters at 5\,K are given in Table~\ref{tab:refinement} and show a good agreement with the earlier publication~\cite{gandara12}.
\subsection{Fourier transform infrared (FT-IR) spectra}
Fourier transform infrared (FT-IR) spectra were measured on a Bruker Equinox 55 FT-IR spectrometer in the range of 4000-400 cm$^{-1}$ with a PLATINUM ATR unit and a KBr beam splitter.
\subsection{SEM micrographs}
The SEM micrographs were taken on a Zeiss Crossbeam 550 Gemini 2 FIB-SEM.
\subsection{Synthesis of Mn(ta)$_{2}$}
Mn(NO$_{3}$)$_{2}$$\cdot$H$_{2}$O (800 mg, 4.06 mmol) was dissolved in 40 mL of DEF, 1$H$-1,2,3-triazole (0.6 mL, 715.2 mg, 10.35 mmol) was added into the 200 ml ACE pressure tube and capped with a silicon O-ring at the front seal of the PTFE bushing. The mixture was heated to 120$^\circ$C in a heating block for 3 days. The product was filtered and washed successively with 3 times 10 mL of DEF and 3 times 10 mL of MeOH and afterwards soaked in 10 mL of MeOH over 3 days, exchanging the MeOH 3 times. Drying overnight under vacuum at RT afforded the phase pure product as a white powder (419 mg, 55\%). FT-IR (ATR) 4000-400 cm$^{-1}$: 3143 (w), 1722 (w), 1653 (w), 1457 (w), 1420 (w), 1226 (w), 1202 (w), 1178 (m), 1096 (s), 993 (w), 974 (m), 794 (s), 721 (w).
\subsection{Synthesis of Zn(ta)$_{2}$}
ZnCl$_{2}$ (500 mg, 3.67 mmol) was dissolved in a mixture of DMF (5 mL), EtOH (5 mL), H$_{2}$O (14 mL) and NH$_{4}$OH (6 mL, 25\%). Dropwise addition of the 1$H$-1,2,3-triazole (0.625 mL, 745 mg, 10.79 mmol) directly resulted in a white precipitate and the suspension was stirred for 24 h. The product was filtered and washed successively with 3 times 10 mL of DEF and 3 times 10 mL of MeOH and afterwards soaked in 10 mL of MeOH over 3 days, exchanging the MeOH 3 times. Drying overnight under vacuum at RT afforded the phase pure product as a white powder (396 mg, 54\%). FT-IR (ATR) 4000-400 cm$^{-1}$: 3145 (w), 1723 (w), 1654 (w), 1461 (w), 1423 (w), 1236 (w), 1213 (w), 1189 (m), 1106 (s), 996 (w), 976 (m), 796 (s), 722 (w).
\begin{table}
	\caption{\label{tab:refinement}
		Refined atomic positions and displacement parameters ($U_{\rm iso}$, in $10^{-2}$\,\r A$^2$) of Mn(ta)$_2$ at 5\,K. The position of hydrogen was not refined. The cubic lattice parameter is $a=18.12711(2)$\,\r A, and the space group is $Fd\bar 3m$ (origin choice 2). 
	}
	\begin{minipage}{0.7\textwidth}
		\begin{ruledtabular}
			\begin{tabular}{cccccc}
				atom & site & $x/a$ & $y/b$ & $z/c$ & $U_{\rm iso}$ \\
				Mn1 & $16d$ & 0            & $\frac12$ & 0           & 0.47(2) \\
				Mn2 & $8b$  & $\frac78$    & $\frac38$ & $\frac78$   & 0.36(2) \\
				N1  & $96g$ & 0.0406(1)    & 0.4180(1) & $y+\frac12$ & 0.30(4) \\
				N2  & $48f$ & $-0.0003(1)$ & $\frac38$ & $\frac78$   & 0.43(5) \\
				C   & $96g$ & 0.1111(1)    & 0.4010(1) & $y+\frac12$ & 0.86(6) \\
				H   & $96g$ & 0.1552       & 0.4233    & $y+\frac12$ & 0.5 \\
			\end{tabular}
		\end{ruledtabular}
	\end{minipage}
\end{table}
\begin{figure}
	\centering\includegraphics[width=8cm]{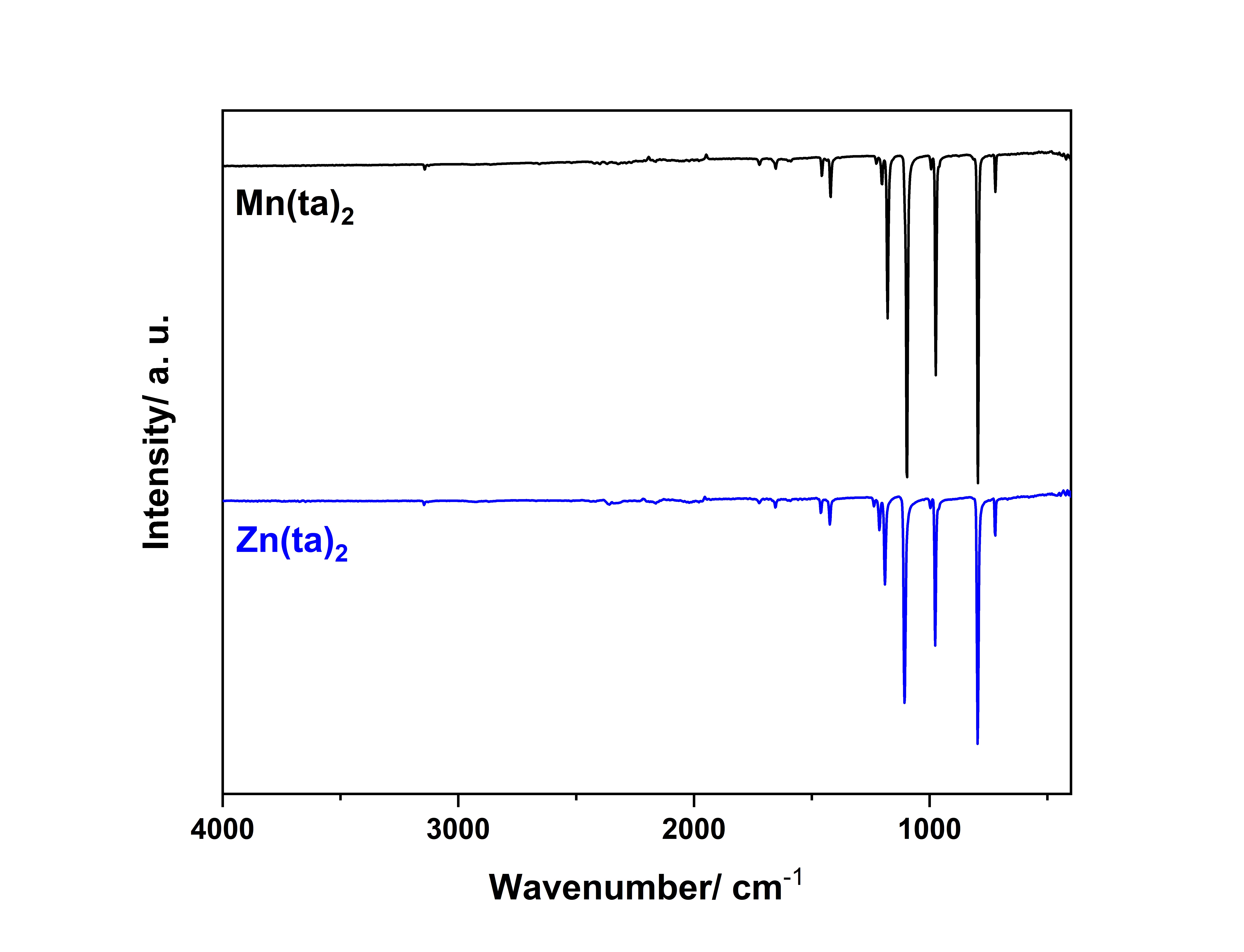}
	\caption{FT-IR(ATR) spectra of Mn(ta)$_{2}$ (black) and Zn(ta)$_{2}$ (blue)}
	\label{fig:FTIR}
\end{figure}
\begin{figure}[h]
	\centering
	\includegraphics[width=7.5cm]{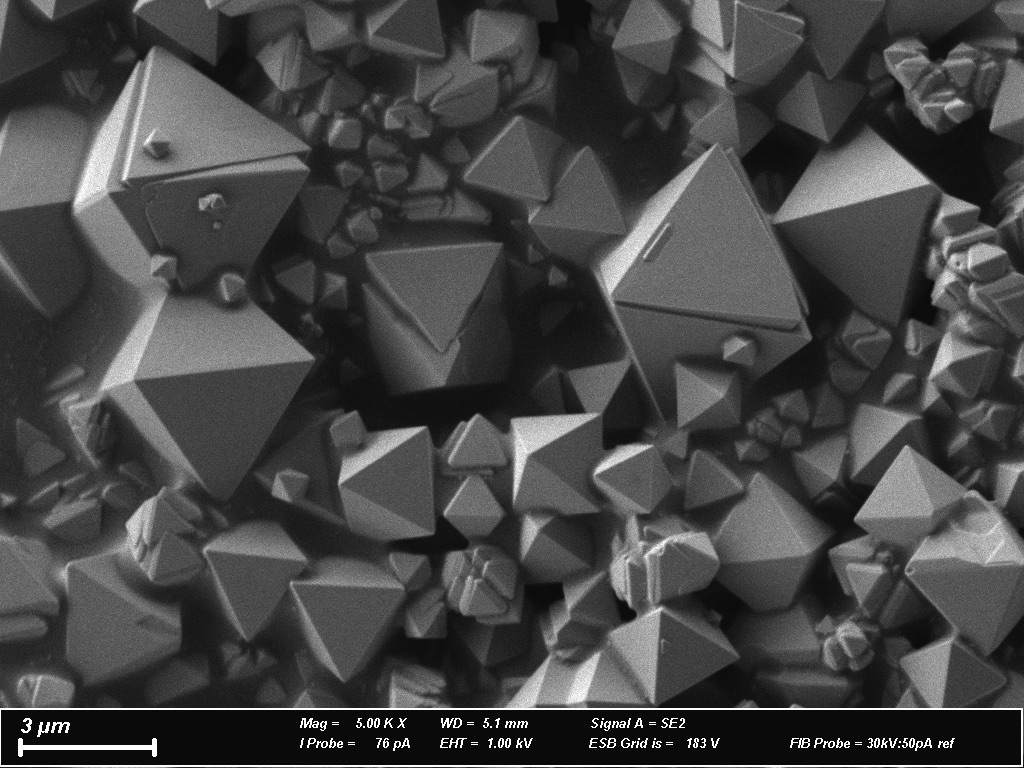}
	\includegraphics[width=7.5cm]{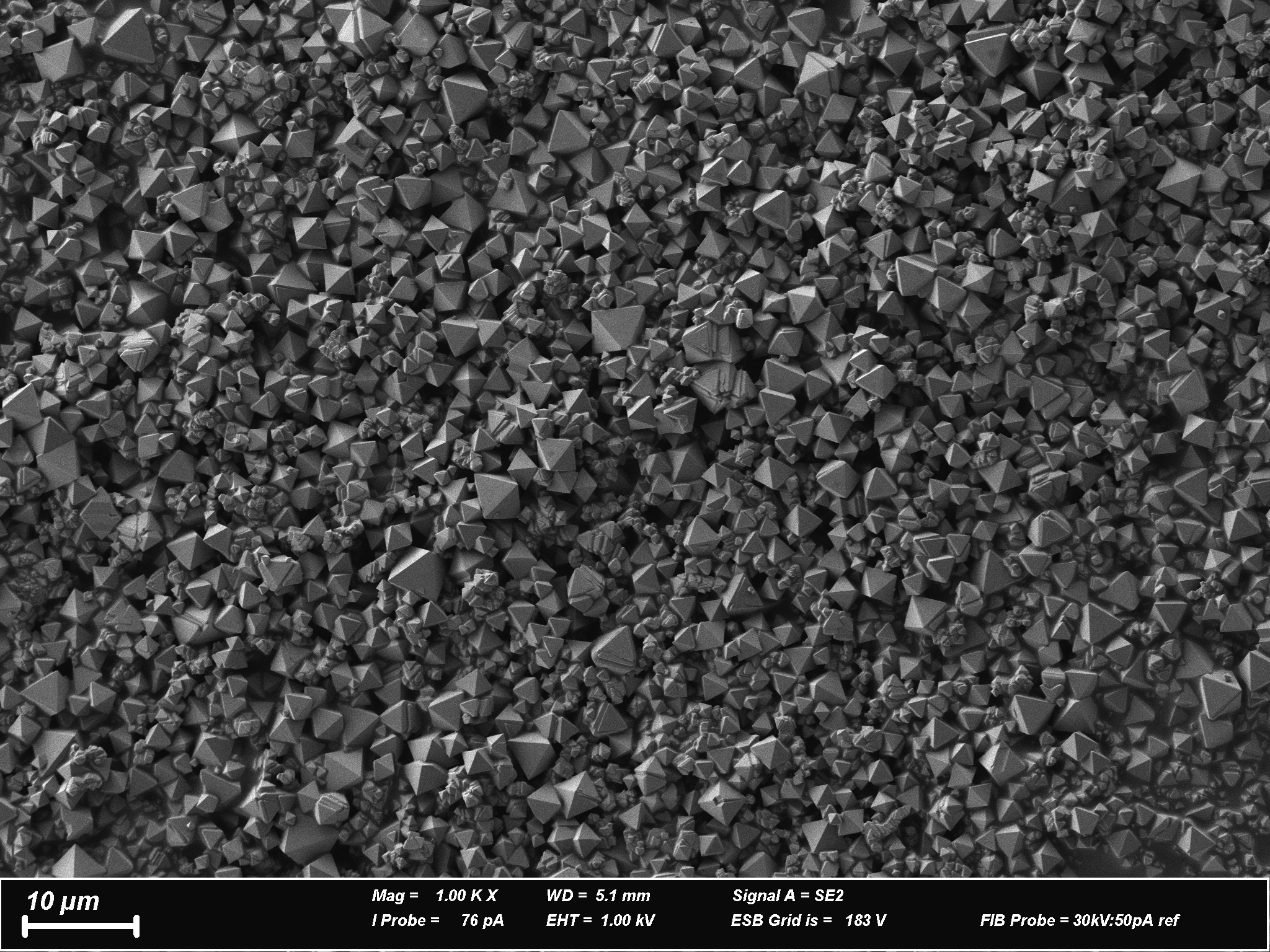}
	\includegraphics[width=7.5cm]{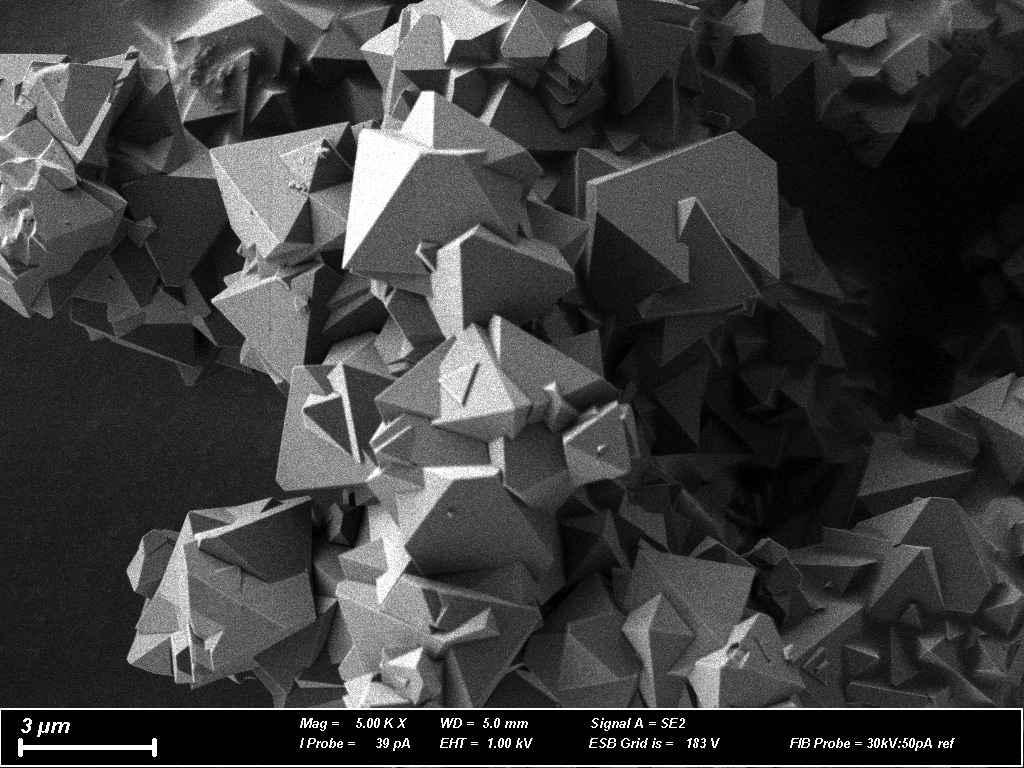}
	\includegraphics[width=7.5cm]{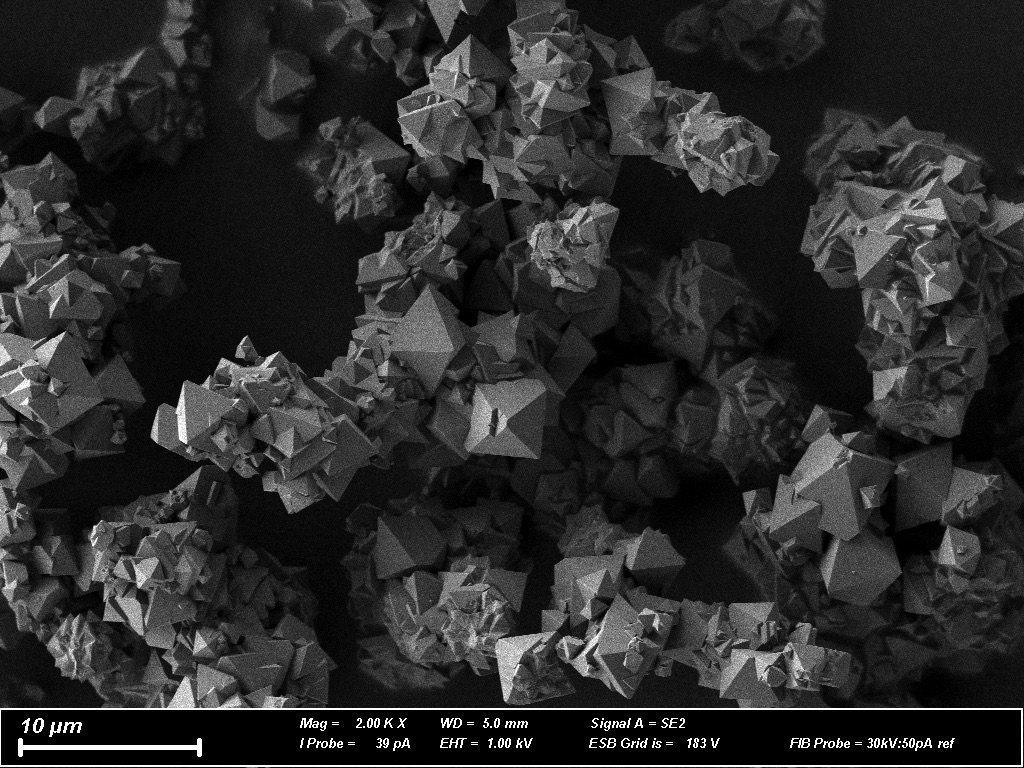}
	\caption{SEM micrographs of Mn(ta)$_{2}$ (\textit{top panels}) and Zn(ta)$_{2}$ (\textit{bottom panels}).}
	\label{fig:SEM}
\end{figure}
\clearpage
\section{Thermodynamic Measurements}
\subsection{Magnetic measurements}
\begin{figure}[h]
	\includegraphics[width=0.9\textwidth]{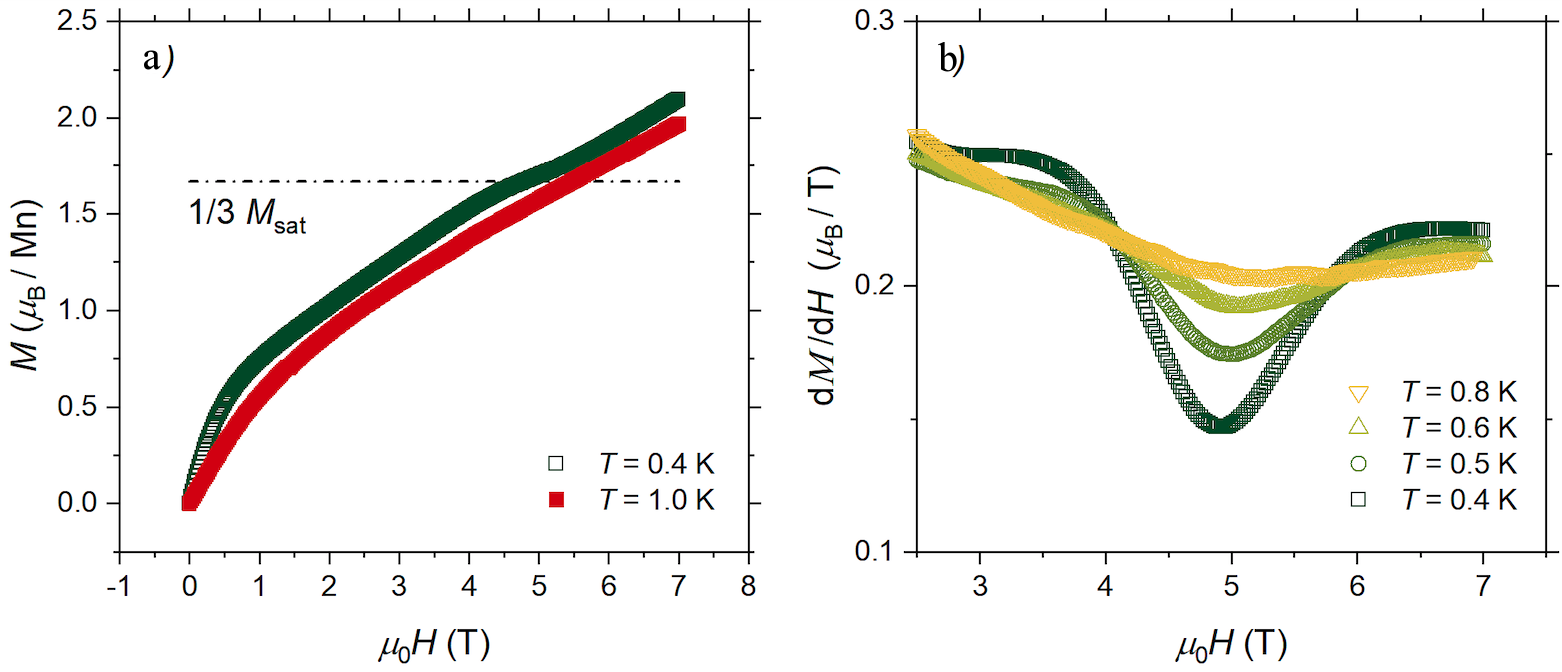}
	\caption{a) Isothermal magnetisation $M$ vs external field $\mu_0 H$ showing a kink at $1/3$ saturated magnetisation ($g_{S}\,S / 3= 1.67 \mu_{B}$) for $T = 0.4 \: \mathrm{K}$. b) The field derivative of $M$ shows that the kink disappears above 0.8 K.
		\label{fig:HCmag}}
\end{figure}

A Quantum Design MPMS3 magnetometer equipped with an iQuantum He3 option was employed to measure the magnetisation at temperatures of $T = 0.4 - 400$~K in applied fields $\mu_{0} H = 0.02$~T below 20 K, and 7 T above 20 K. 20K was chosen as a good compromise between the increasing noise/signal ratio of the 20mT data at higher temperatures, and deviations from the linear regime at 7T at lower temperatures. fig. 2b of the main text confirms that we are in the linear regime up to 7T. The powder was mounted using a plastic capsule that was also measured empty in order to determine the background contribution for $T > 2$ K. The error of the absolute value of the measured magnetic moment in the He3 region ($T < 2$ K) is larger than the one at higher temperatures ($T > 2$K). Therefore, a small offset ($\sim$5\% of the absolute value, which is smaller than the size of the data points in fig. 2b was subtracted from the He3 data such that it fits the susceptibility obtained at $T = 2$ K.\\
In the ordered phase, the magnetisation curve at $T=0.4$~K deviates from the linear behaviour near $1/3$ of the saturated magnetisation (fig. \ref{fig:HCmag}.a), suggesting a potential ferrimagnetic order. This feature becomes less prominent above $T_c$ and vanishes at 0.8 K (fig. \ref{fig:HCmag}.b).
\subsection{Specific-heat measurements}
\begin{figure}[b]
	\includegraphics[width=0.6\textwidth]{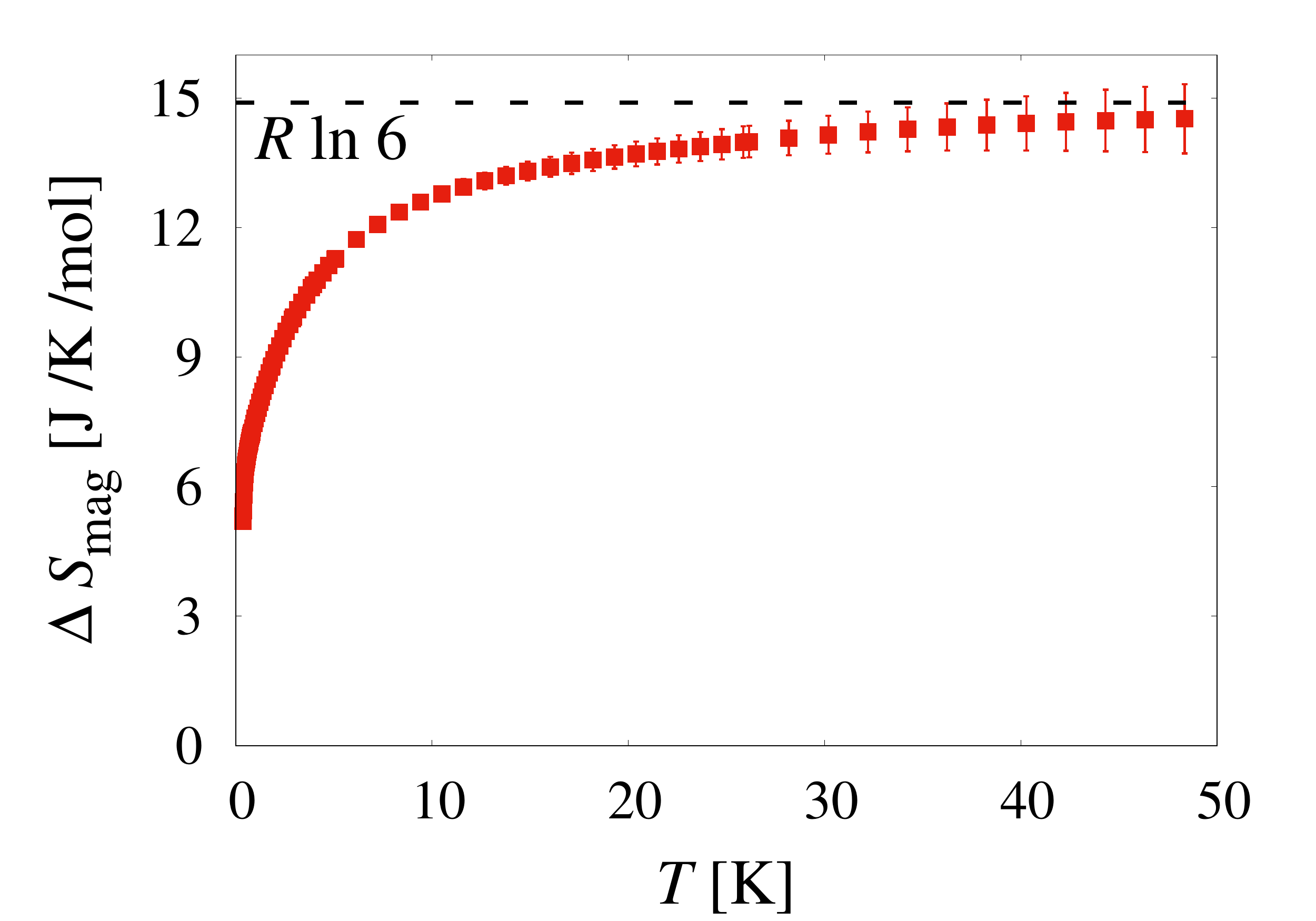}
	\caption{Temperature-dependent magnetic contribution to the entropy of [Mn(ta)$_{2}$].
		\label{fig:entropy}}
\end{figure}

Specific heat was measured on a pressed pellet by a relaxation method using Quantum Design PPMS with the He3 insert. In order to determine non-magnetic contributions, the specific heat of the non-magnetic analogue [Zn(ta)$_{2}$] was measured in the full temperature range. For $T < 1.8$ K, the magnetic contribution was found to amount to more than 99.7\% of the absolute value and the non-magnetic host is negligible. For increasing temperatures to $T > 2$\,K the latter becomes more important and has to be subtracted to get reliable values for $C_{\rm mag}$. To this extent, the temperature axis of the [Zn(ta)$_{2}$] for $C(T)$ had to be re-scaled to account for a different Debye temperature; a factor of 0.91 was applied such that specific heat of [Mn(ta)$_{2}$]and [Zn(ta)$_{2}$] approach identical values at $T \approx 50$\,K.\\
fig. \ref{fig:entropy} shows the temperature-dependent, magnetic contribution to the change of magnetic entropy, $\Delta S_{\mathrm{mag}}$.
Since contributions for $T < 0.36$\,K are not accessible using the available equipment, the entropy curve was shifted such that the full entropy of an $S = 5/2$ system is obtained at high temperatures. Accordingly, we estimate a low temperature contribution of $S = 5.3$ J\,mol$^{-1}$K$^{-1}$.\\
$C_{\rm mag}(T)$ measured in applied magnetic fields is shown in fig. \ref{fig:dipolar_exp}b for temperatures $T < 0.8$\,K. 
The anomaly broadens in applied field and slightly shifts to higher temperatures in accordance with the proposed ferrimagnetic ordering. The entropy was found to increase above and below $T_c$. For larger applied fields, an increasing shift of entropy to higher temperatures is observed.   
\clearpage
\section{Density-Functional Calculations}
Density-functional calculations were performed in the FPLO code \cite{Kopernik99} using the experimental crystal structure of [Mn(ta)$_{2}$] \cite{He17}. Magnetic exchange couplings defined in Eq. 1 of the main text with $S = 5/2$ of Mn$^{2+}$ were obtained by a mapping procedure \cite{Xiang11}, detailed below, with total energies of individual spin configurations converged on a $k-$mesh with 64 points in the first Brillouin zone. The Perdew-Burke-Ernzerhof exchange-correlation potential \cite{Perdew96} was used, and correlations effects in the Mn 3d shell were taken into account of the mean-field DFT$+U$ level with the on-site Coulomb repulsion parameter $U_{d} = 5$~eV and Hund's coupling $J_{H} = 1$~eV. This choice of parameters not only led to a good agreement with the experiment, but also showed consistency with the previous ab initio calculations for other Mn$^{2+}$ frustrated magnets \cite{Nath14}.
\subsection{Exchange Couplings}
\begin{figure}[h]
	\includegraphics[width=0.7\textwidth]{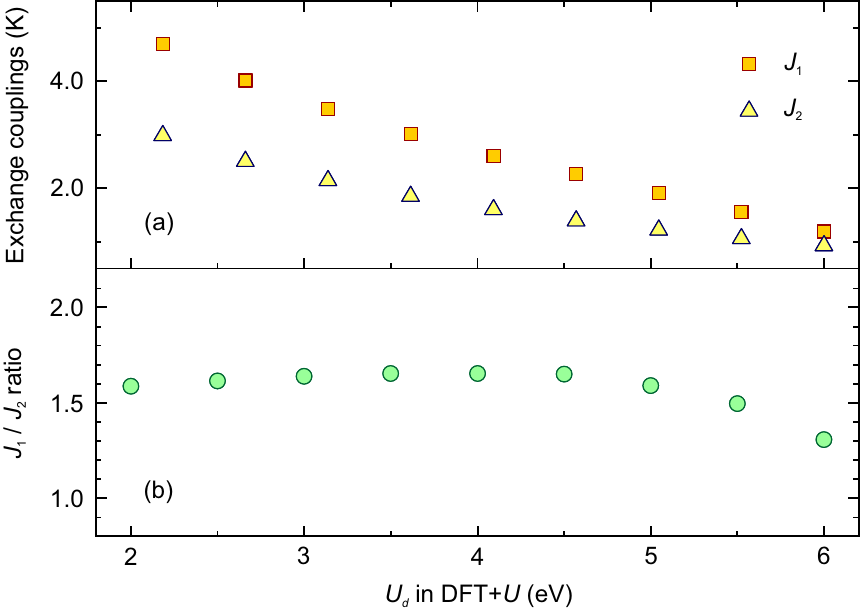}
	\caption{DFT results for the exchange couplings: (a) $J_1$ and $J_2$ decrease with increasing the $U_d$ parameter of DFT+$U$; (b) the $\gamma=J_1/J_2$ ratio does not change significantly.
		\label{J-vs-U}}
\end{figure}

Exchange couplings were evaluated by a mapping procedure. To this end, we compute total energies of several collinear spin configurations within the primitive unit cell of Mn(ta)$_2$. The primitive cell contains six Mn atoms, two of them belonging to the Mn(1) site and the remaining four to the Mn(2) site. We have used five spin configurations, as follows: 
\begin{equation*}
	1: (++++++)\qquad 2: (-+++++)\qquad 3: (++-+++)\qquad 4: (--++++)\qquad 5: (++--++)
\end{equation*}
where the first two atoms are Mn(1) and the remaining four are Mn(2). Then, 
\begin{equation}
	J_1=\frac{E_1-2E_2+E_4}{4S^2},\qquad J_2=\frac{E_1-E_2-E_3+E_5}{4S^2}.
\end{equation}
The resulting exchange couplings depend on the $U_d$ parameter of DFT+$U$ (fig. \ref{J-vs-U}). Both $J_1$ and $J_2$ become smaller when $U_d$ is increased and roughly follow the $1/U_d$ dependence expected for antiferromagnetic superexchange. On the other hand, the ratio $\gamma=J_1/J_2$ is only weakly affected by the choice of $U_d$ and falls in the range of $1.30-1.65$ regardless of the $U_d$ value.
\section{Theory}
\subsection{Lattice}
The centred pyrochlore lattice is defined by the positions
\begin{equation}
	\mathbf{r}_{i,\mu} = \mathbf{R}_{i} + \bm{\delta}_{\mu},
\end{equation}
with a face-centred cubic Bravais lattice $\mathbf{R}_i = n_1\mathbf{a}_1 + n_2\mathbf{a}_2 + n_3\mathbf{a}_3$, $n_i \in \mathbb{Z}$, and lattice vectors $\mathbf{a}_1 = \frac{1}{2}(1,1,0)$, $\mathbf{a}_2 = \frac{1}{2}(0,1,1)$, $\mathbf{a}_3 = \frac{1}{2}(1,0,1)$.
The six sublattices have basis vectors
\begin{eqnarray}
	&\bm{\delta}_a = \mathbf{0},
	\:
	\bm{\delta}_b = \frac{1}{4}
	\begin{pmatrix}
		1\\1\\1
	\end{pmatrix},
	\:
	\bm{\delta}_0 = \frac{1}{8}
	\begin{pmatrix}
		1\\1\\1
	\end{pmatrix},
	\nonumber \\ 
	&\bm{\delta}_1 = \frac{1}{8}
	\begin{pmatrix}
		1\\-1\\-1
	\end{pmatrix},
	\:
	\bm{\delta}_2 = \frac{1}{8}
	\begin{pmatrix}
		-1\\1\\-1
	\end{pmatrix},
	\:
	\bm{\delta}_3 = \frac{1}{8}
	\begin{pmatrix}
		-1\\-1\\1
	\end{pmatrix}.
	\label{eq:deltas}
\end{eqnarray}
All quantities are given in units where the side length of the conventional fcc unit cell $a = 1$. Sites at the centres of tetrahedra are labelled by $\mu \in \{a,b\}$ and those at the corners by $\mu \in \{0\dots 3\}$.
\subsection{Flat Bands}
For $\gamma < 4$ the CPy model hosts a four-fold degenerate flat band in its mean-field spectrum.
This can be seen by adapting the argument of refs.\cite{katsura10,essafi17} to the centred pyrochlore geometry.
The Hamiltonian (eq. 1 of the main text) can be rewritten for $\mathbf{H} = 0$ in terms of an $\frac{N}{3} \times N$ rectangular matrix, $\mathrm{A}_{t,n}$,
\begin{equation}
	H = \frac{J_2}{2}\sum_{t=1}^{N/3} \sum_{n,m = 1}^N \mathrm{A}_{t,n} \mathrm{A}_{t,m} \mathbf{S}_n\cdot \mathbf{S}_m + \mathrm{const},
\end{equation}
where $N$ is the total number of spins.
The elements of $\mathrm{A}$ are given by
\begin{equation}
	\mathrm{A}_{t,n} = 
	\begin{cases}
		1, \qquad \text{if $n$ $\in$ corners of $t$}\\
		\gamma, \qquad \text{if $n$ $\in$ centre of $t$}\\
		0, \qquad \text{otherwise}
	\end{cases}.
\end{equation}
The labels $n,m$ enumerate all sites of the lattice, whereas $t$ enumerates the tetrahedra.
Any eigenstate of $\mathrm{A}$ with eigenvalue $0$, is a ground state of the system.
Therefore the number of ground states is given by the dimension of the null space of the matrix $\mathrm{A}$.
Since
\begin{equation}
	\mathrm{rank}(\mathrm{A}) \leq \frac{N}{3},
\end{equation}
the dimension of the null space,
\begin{equation}
	\mathrm{Nullity}(\mathrm{A}) \geq N - \frac{N}{3} = \frac{2N}{3},
\end{equation}
by the rank-nullity theorem (see for example refs. \cite{mathai_2017,katznelson_2007}).
The dimension of a band in reciprocal space is $\frac{N}{6}$, so assuming the ground states correspond to flat bands, $4$ out of $6$ bands of the mean-field energy spectrum of the CPy model must be flat and belong to the ground state.
Applying the generalized Luttinger-Tisza method \cite{lyons60} to the CPy model, one also finds that, for $\gamma < 4$, the mean-field energy spectrum consists of a four-fold degenerate flat band with a gap to excitations \cite{Nutakki2023}.
\subsection{Monte Carlo Simulations}
Simulations were performed for cubic $L\times L\times L$ systems with periodic boundary conditions. 
$L$ is the number of conventional cubic unit cells in $x$, $y$ and $z$ directions, giving a total of $N = 24L^3$ spins with all simulations performed up to at least a system size of $L=10$. 
We used single-spin heat-bath \cite{creutz80,janke08,greitemann19} and over-relaxation \cite{brown87,creutz87,landau09,greitemann19} updates, with one Monte Carlo (MC) step made up of a sweep of both updates through the entire lattice. 
Simulations were typically performed using a minimum of $10^5$ thermalization and measurement steps, increasing up to a maximum of $10^7$ thermalization and measurement steps to access lower temperatures. 
In each MC measurement step, we calculated the magnetisation,
\begin{equation}
	\mathbf{m} = \frac{1}{N} \bigg\langle \sum_i \mathbf{S}_i \bigg\rangle,
\end{equation}
where $\mathbf{m} = (m_x,m_y,m_z)^T$, magnetic susceptibility (per site) along a single axis
\begin{equation}
	\chi_z = \frac{N}{T} \bigg( \langle m_z^2 \rangle - \langle m_z \rangle ^2 \bigg),
\end{equation}
and specific heat 
\begin{equation}
	c = \frac{1}{NT^2}\bigg(\langle E^2 \rangle - \langle E \rangle^2 \bigg)
	\label{eq:c}
\end{equation}
where $E$ is the energy, for system sizes up to $L = 12$. 
For comparison to experiments, figs. 2a and 2b of the main text, the magnetisation $m_{z}$ and susceptibility $\chi_{z}$ were scaled by $\left(1+1/S\right)=1.4$ to account for the difference between classical continuous spins and quantum $S=5/2$ spins.
We also computed the structure factors defined below every 100 MC measurement steps, for system sizes up to $L = 10$.
Firstly, the static spin structure factor,
\begin{equation}
	\mathcal{S}_s(\mathbf{q}) = \frac{1}{N_s} \sum_{j,k=1}^{N_s} e^{i\mathbf{q}\cdot (\mathbf{r}_j-\mathbf{r}_k)} \langle \mathbf{S}_j \cdot \mathbf{S}_k \rangle,
\end{equation}
where $\mathbf{r}_{j}(\mathbf{r}_{k})$ is the vector position of site $j(k)$, with the sums over both $j,k$ either running over $s = \text{all}$ lattice sites or only $s = \text{corner}$ sites, depending on the structure factor we wanted to compute.
Secondly, to investigate Coulomb physics in the CPy model, we compute
\begin{equation}
	\mathcal{B}^{\alpha}_{\mu \nu}(\mathbf{q}) = \frac{1}{N_{\mathrm{centres}}} \sum_{n,m=1}^{N_{\mathrm{centres}}} e^{i\mathbf{q}\cdot (\mathbf{r}_m-\mathbf{r}_n)} \langle B^{\alpha}_{\mu}(\mathbf{r}_n)B^{\alpha}_{\nu}(\mathbf{r}_m) \rangle,
\end{equation}
where the fields
\begin{equation}
	\mathbf{B}^{\alpha}(\mathbf{r}_n) = (B^{\alpha}_x(\mathbf{r}_n),B^{\alpha}_y(\mathbf{r}_n),B^{\alpha}_z(\mathbf{r}_n))^T = \sum_{i=1}^4 S^{\alpha}(\mathbf{r}_n \pm \delta_i) \hat{\bm{\delta_i}},
	\label{eq:field_def}
\end{equation}
are defined at the centres of the tetrahedra, $\mathbf{r}_n$, with $\pm = +(-)$ when $n$ is an a(b) tetrahedron, $\alpha = {x,y,z}$ labels the spin components and $\hat{\bm{\delta_i}} = \frac{\bm{\delta_i}}{\lvert \bm{\delta_i}\rvert}$, see eq. \ref{eq:deltas}.
This is the usual definition (up to a prefactor) as described in refs. \cite{isakov04,henley05,henley10} for the pyrochlore, where the ground state constraint leads to a zero-divergence condition for this field.
\subsection{Analysis of Structure Factors}
\begin{figure}[h]
	\includegraphics[width=\textwidth]{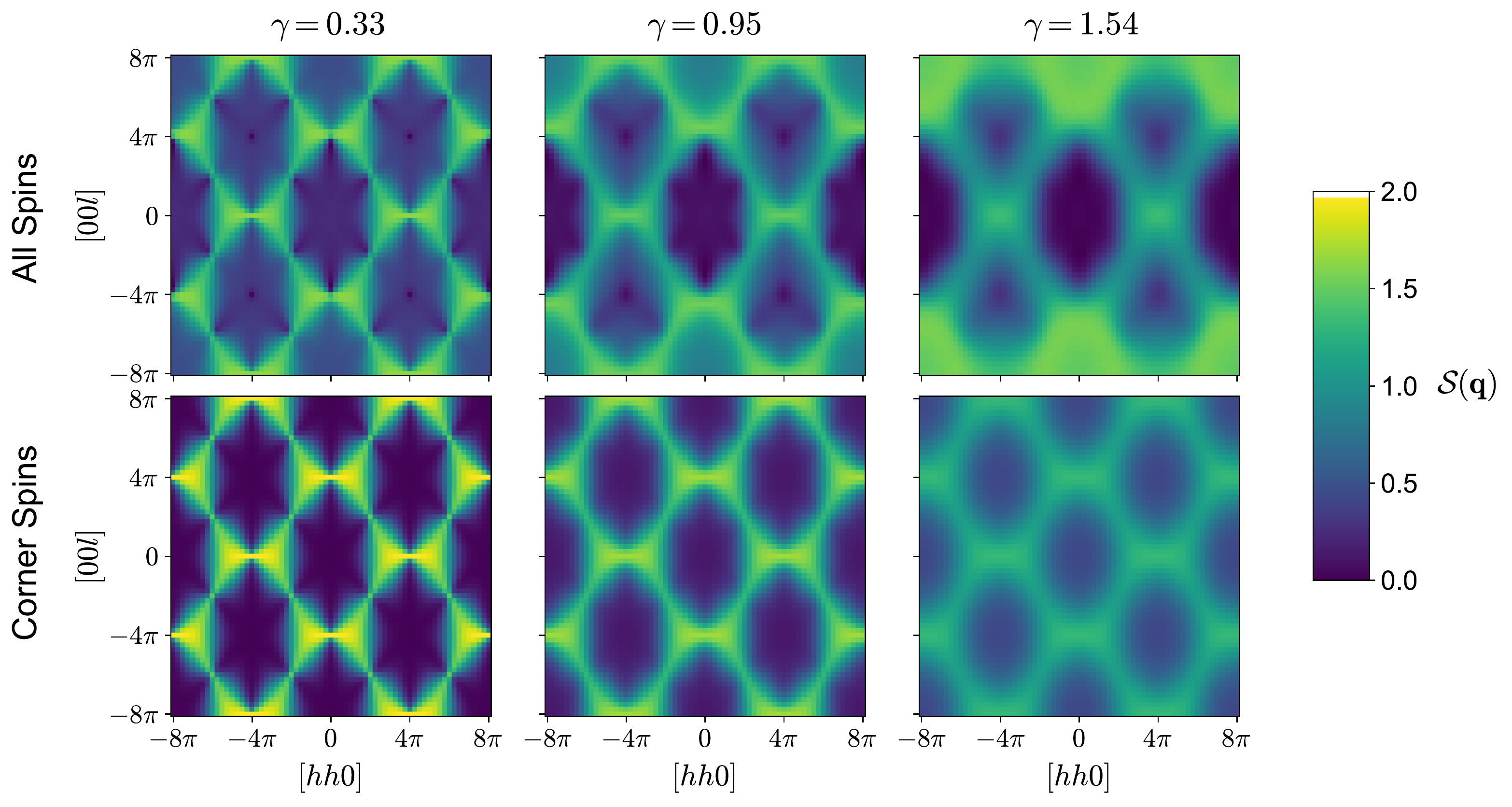}
	\caption{Spin structure factors computed from MC simulations at $\frac{T}{J_1S^2} = 0.005$ and various $\gamma$, including all spins (top panels) and only corner spins (bottom panels). Pinch points broaden and the relative contribution of other features increases with $\gamma$.}
	\label{fig:sf}
\end{figure}

\begin{figure}
	\includegraphics[width=0.8\textwidth]{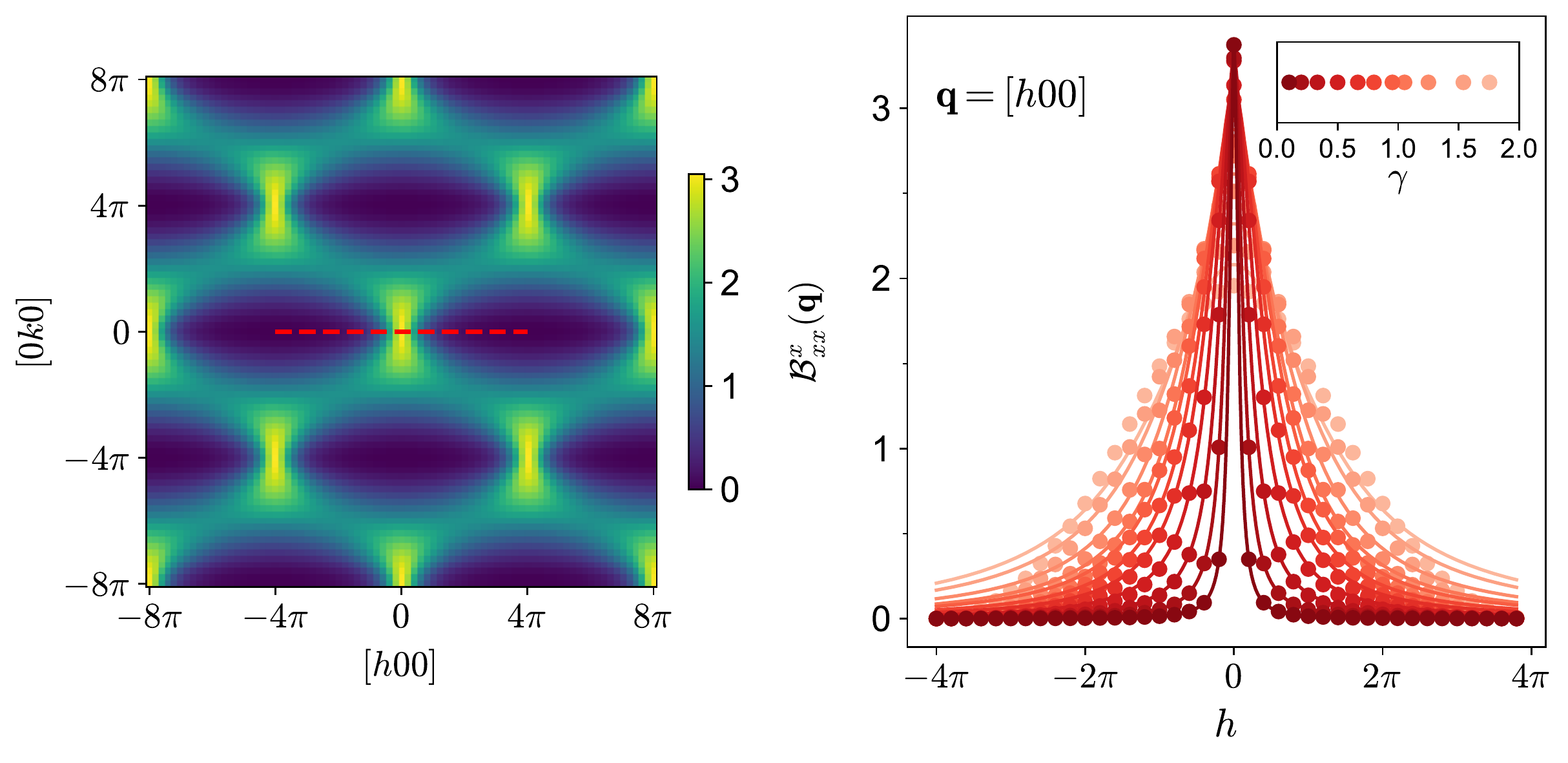}
	\caption{Extracting $\kappa$ from the $\mathcal{B}^x_{xx}$ structure factor computed from MC simulations at $\frac{T}{J_1S^2} = 0.005$ for various $\gamma$. Left panel, $\mathcal{B}^x_{xx}(\mathbf{q})$ for $\gamma = 0.66$, with the red line showing the cut along which a Lorentzian is fitted. Right panel, $\mathcal{B}^x_{xx}(\mathbf{q} = (h,0,0))$ (circles) and Lorentzian fits (lines).}
	\label{fig:B_xx}
\end{figure}

Spin structure factors computed from MC simulations in the CPy spin liquid state at low temperature are shown in fig. \ref{fig:sf}.
At small $\gamma$, the dominant feature of the structure factor is the characteristic bow ties of the pyrochlore, with central spins adding only a diffuse contribution.
As $\gamma$ increases the bow ties broaden and weight in the structure factor of corner spins is transferred to a diffuse and approximately constant background. 
In the structure factor of all spins, correlations between centre-centre and centre-corner spins begin to wash out the bow ties.\\
The broadening of the bow ties can be attributed to breaking of the divergence-free condition for the field $\mathbf{B}^{\alpha}$, which results in entropic Debye screening of the resulting ``charges" \cite{henley10}.
Thus the previously sharp pinch points acquire a finite width with the form
\begin{equation}
		\mathcal{B}^{\alpha}_{\mu \nu}(\mathbf{q}) \propto \delta_{\mu \nu} - \frac{q_{\mu}q_{\nu}}{q^2 + \kappa^2},
\end{equation}
where $\kappa$ parameterizes the width.
This corresponds in real space to screened correlations of the form $\frac{e^{-\kappa r}}{r^3}$, where $1/\kappa$ is the screening length.
We fit $\mathcal{B}^{x}_{xx}(\mathbf{q})$ obtained from MC simulations to the Lorentzian
\begin{equation}
	\mathcal{B}^{x}_{xx}(q_x,q_y=0,q_z=0) = \frac{A}{q^2_x+\kappa^2}
\end{equation}
for various $\gamma$ with $A$ and $\kappa$ fitting parameters, as shown in fig. \ref{fig:B_xx}.
We are thus able to plot $\kappa$ against $\gamma$, fig. 3c of the main text, where we find that $\kappa \propto \gamma$ up to $\gamma \approx 1.25$, consistent with a description of the spin liquid as a dilute charge fluid where the charge strength is parameterized by $\gamma$.

\subsection{Specific Heat}
\subsubsection{Exact Diagonalization}
Full diagonalization of the Hamiltonian was performed on a single tetrahedron consisting of five $S = \frac{5}{2}$ quantum spins with open boundary conditions. The specific heat, see Eq. \ref{eq:c}, was calculated from the resulting energy spectrum at various temperatures. Results for various $\gamma$ were compared to experiment using the energy scale, $J_1$, and a rescaling factor for $c$ as free parameters. This is plotted in the main text in Fig. 2c.
\subsubsection*{Monte Carlo}
\begin{figure}
	\includegraphics[width=0.4\textwidth]{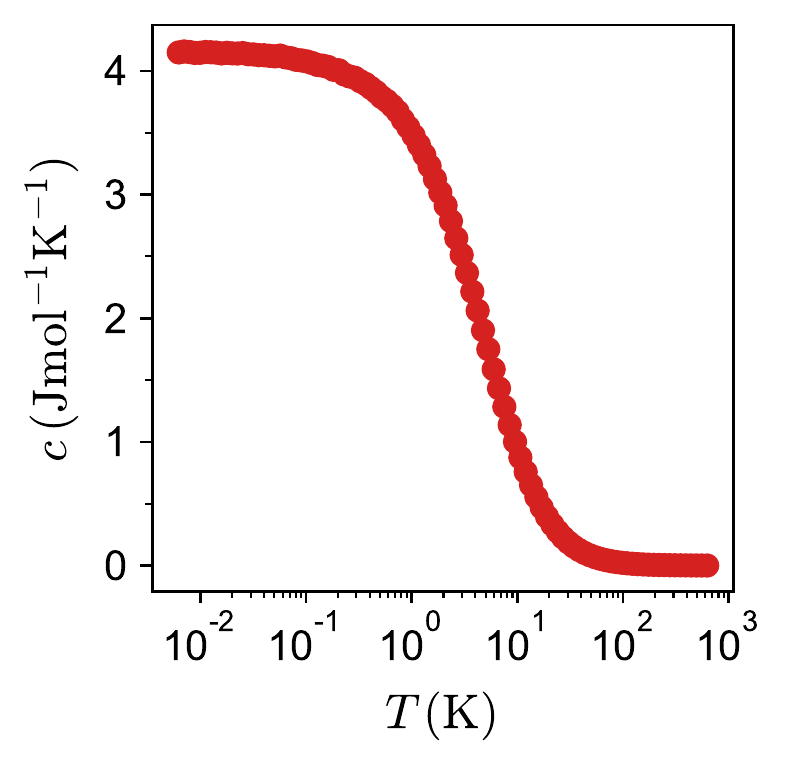}
	\caption{Specific heat from MC simulations without dipolar interactions for $J_1^{\mathrm{MC}} = 2.0, \gamma^{\mathrm{MC}} =  1.51$.}
	\label{fig:cT_mc}
\end{figure}

The specific heat obtained from MC simulations for the optimal Mn(ta)$_2$ parameters is shown in fig. \ref{fig:cT_mc}.
The spin liquid is characterized by a plateau in the specific heat at $c = \frac{1}{2} k_B N_A = 8.31 \: \mathrm{Jmol^{-1}K^{-1}}$.
The specific heat reaches half of this value at $T \approx 4 \: \mathrm{K}$; this temperature may be used as an estimate for the crossover between the paramagnet and spin liquid. 
Furthermore, at this temperature the spin structure factor qualitatively matches that observed at lower temperatures.
\subsection{Dipolar Interactions}
To include the effect of dipolar interactions, we extend the CPy Hamiltonian to the form
\begin{equation}
	H_D = J_1 \sum_{\langle ij \rangle} \mathbf{S}_i \cdot \mathbf{S}_j + J_2 \sum_{\langle \langle ij \rangle \rangle} \mathbf{S}_i \cdot \mathbf{S}_j + D \sum_{(ij):r_{ij} < R} \bigg( \frac{\mathbf{S}_i \cdot \mathbf{S}_j}{r_{ij}^3} - 3\frac{(\mathbf{S}_i\cdot\mathbf{r}_{ij})(\mathbf{S}_j \cdot \mathbf{r}_{ij})}{r_{ij}^5} \bigg)
	\label{eq:dip_ham}
\end{equation}
where the displacement vector $\mathbf{r}_{ij} = \mathbf{r}_j - \mathbf{r}_i$ and the dipolar term sums over all $ij$ pairs within the cutoff radius $R$.
The dipolar interaction strength is given by $D = \frac{g_{\mathrm{exp}}^2 \mu_B^2 \mu_0}{4\pi k_B a^3_{\mathrm{exp}}} = 0.043r^3_{\mathrm{nn}} \: \mathrm{K}$, where $g_{\mathrm{exp}}$ and $a_{\mathrm{exp}}$ are the experimentally determined g-factor and cubic unit cell length respectively.
$r_{\mathrm{nn}}$ is the nearest neighbour distance in units where $a = 1$.
For $S = 5/2$ spins, the nearest neighbour dipolar interaction strength is $D_{\mathrm{nn}} = 0.27 \: \mathrm{K}$.
In what follows we fix the exchange couplings $J_1^{\mathrm{MC}} = 2 \: \mathrm{K}$, $\gamma^{\mathrm{MC}} = 1.51$.

\begin{figure}
	\includegraphics[width=\textwidth]{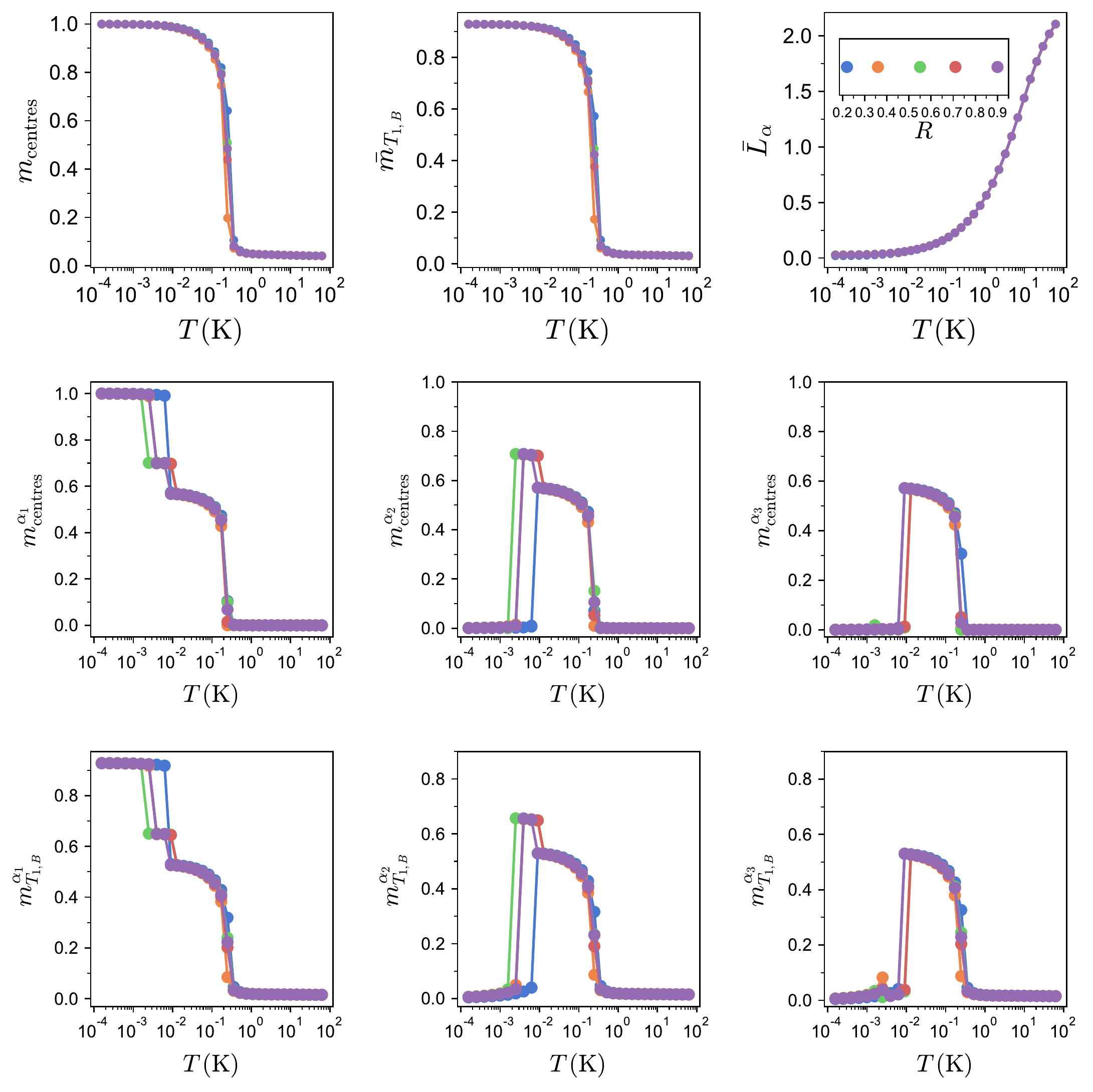}
	\caption{MC results for the quantities which capture the ordering found in the model described by $H_D$ with experimentally relevant parameters. Results shown are for $L = 4$ and cut-off radii (in units of $a = 1$) indicated in the inset. \textbf{a-c} Show the absolute values of the magnetizations and local constraint,  \textbf{d-e} the components of $\mathbf{m}_{\mathrm{centres}}$, \textbf{g-i} the components of $\bar{\mathbf{m}}_{T_{1,B}}$. We checked that this ordering is robust to the inclusion of longer range interactions (up to $R = 4$) and upon increasing the system size (up to $L = 8 = $ over 12000 spins).}
	\begin{textblock}{1}(-0.2,-9.6)
		\textbf{a.}
	\end{textblock}
	\begin{textblock}{1}(3.8,-9.6)
		\textbf{b.}
	\end{textblock}
	\begin{textblock}{1}(7.9,-9.6)
		\textbf{c.}
	\end{textblock}
	\begin{textblock}{1}(-0.2,-6.8)
		\textbf{d.}
	\end{textblock}
	\begin{textblock}{1}(3.8,-6.8)
		\textbf{e.}
	\end{textblock}
	\begin{textblock}{1}(7.9,-6.8)
		\textbf{f.}
	\end{textblock}
	\begin{textblock}{1}(-0.2,-3.8)
		\textbf{g.}
	\end{textblock}
	\begin{textblock}{1}(3.8,-3.8)
		\textbf{h.}
	\end{textblock}
	\begin{textblock}{1}(7.9,-3.8)
		\textbf{i.}
	\end{textblock}
	\label{fig:dipolar_order}
\end{figure}

MC simulations find that the inclusion of dipolar interactions induces ordering at $T = 0.25 \: \mathrm{K}$.
This order is almost entirely captured by the magnetization $\mathbf{m}_{\mathrm{centres}}$ measured only on centre sites, the average local constraint, $\bar{L}_{\alpha} = \frac{1}{N_{\alpha}} \sum_{\alpha} \abs{\mathbf{L}_{\alpha}}$, see eq. 3 of the main text, and the average magnetization corresponding to the $T_{1,B}$ irreducible representation, $\bar{\mathbf{m}}_{T_{1,B}}  = \frac{1}{N_\alpha} \sum_{\alpha} \mathbf{m}_{T_{1,B},\alpha}$ as defined (with the same convention for indexing corner sites) in ref. \cite{Yan2017}.
Fig. \ref{fig:dipolar_order} shows MC results for these quantities for $L = 4$ (= over $1500$ spins) and various cut-off radii.
Looking at the vector components of $\mathbf{m}_{\mathrm{centres}}$ and  $\bar{\mathbf{m}}_{T_{1,B}}$ reveals a delicate transition region, with $\mathbf{m}_{\mathrm{centres}}$ changing orientation from $\frac{1}{\sqrt{3}}(111) \rightarrow \frac{1}{\sqrt{2}}(110) \rightarrow (100)$ as $T$ is lowered (up to permutations of the components). 
The orientation of $\bar{\mathbf{m}}_{T_{1,B}}$ also changes correspondingly.
The low temperature ordered state can be understood in the following way.
For $R = r_{\mathrm{nn}}$, eq. \ref{eq:dip_ham} can be rewritten in the form $H_D = \sum_{\alpha} H_{\alpha} + \mathrm{const}$, where
\begin{equation}
	H_{\alpha} = \frac{J_2}{2}\abs{\mathbf{L}_{\alpha}}^2 + 2\sqrt{2}D_{\mathrm{nn}} \mathbf{S}_{c,\alpha}\cdot \mathbf{m}_{T_{1,B},\alpha}.
\end{equation}
The two terms for $H_{\alpha}$ cannot be simultaneously minimized since spin weight used to satisfy the local constraint reduces the weight available to maximize $m_{T_{1,B},\alpha}$.
Since $J_2 S^2 \gg D_{\mathrm{nn}}$, the dipolar interactions can be thought of as a perturbation to the Heisenberg part of the Hamiltonian. 
Therefore to construct a ground state we should ensure the local constraint (eq. 4 of the main text) is satisfied, which polarizes the corner spins on a tetrahedron antiparallel to the centre spin.
Since $\gamma < 4$, the corner spins retain some finite weight after satisfying the local constraint which can be put into the ${T_{1,B}}$ channel to further minimize the energy.
To verify this picture we propose the single tetrahedron variational state,
\begin{eqnarray}
	&\mathbf{S}_c = (1,0,0), \nonumber\\
	&\mathbf{S}_0 = (-a, b, b), \nonumber\\
	&\mathbf{S}_1 = (-a, -b, -b)\nonumber,\\
	&\mathbf{S}_2 = (-a, -b, b), \nonumber\\
	&\mathbf{S}_3 = (-a, b, -b),\nonumber\\
	&a = \frac{\gamma}{4}, \qquad b=\sqrt{\frac{1}{2}\bigg(1-\frac{\gamma^2}{16}\bigg)},
	\label{eq:var_state}
\end{eqnarray}
which repeating on every single tetrahedron gives a variational state for the entire lattice.
Longer range dipolar interactions may introduce small modifications, but our simulations results indicate that these do not significantly alter the order.
For example, this state very closely reproduces the energies and magnetizations obtained from numerical minimization of eq. \ref{eq:dip_ham} on a single tetrahedron and large-scale MC simulations at low temperature for $R = r_{\mathrm{nnn}}$, as summarized in table II.
Therefore the low temperature ordered state can be described as an unsaturated ferrimagnet with finite magnetic moment along a cubic axis, where on each tetrahedron the corner spins realize a planar antiferromagnet perpendicular to this axis with the remaining spin weight.

\begin{table}
	\begin{ruledtabular}
		\begin{tabular}{llllllll}
			State & $E/N$ & $L_{\alpha}$ & $m_{A_2}$ & $m_E$ & $m_{T_{1,A}}$ & $m_{T_{1,B}}$ & $m_{T_2}$ \\
			\hline
			Numerical minimization & -1.44897 & 0.02924 & 0 & 0 & 0.37005 & 0.92901 & 0\\
			Variational            & -1.44877 & 0 & 0 & 0 & 0.37736 & 0.92607 & 0\\
			MC ($T = 1.6\times 10^{-4} \: \mathrm{K}$) (100) & -1.44900 & 0.02983 & 0.00025 & 0.00044 & 0.36999 & 0.92879 & 0.00103
		\end{tabular}
	\end{ruledtabular}
	\caption{Comparison of energy and corner spin magnetizations corresponding to irreducible representations defined in ref. \cite{Yan2017} with $R=r_{\mathrm {nnn}}$. Numerical minimization is performed for a single tetrahedron, the variational results are computed from the state defined in eq. \ref{eq:var_state} and MC results are averaged over all tetrahedra in the lattice. Note $m_{T_{1,A}}$ is the total magnetization of corner spins on a tetrahedron so captures the finite moment of corner spins in order to ensure $L_{\alpha} \approx 0$. In the single tetrahedron calculations we always find $\mathbf{S}_c = (1,0,0)$ (up to permutation of spin components) which corresponds to $\mathbf{m}_{\mathrm{centres}} = (1,0,0)$ on the full lattice.} 
\end{table}

\begin{figure}
	\begin{minipage}{0.66\textwidth}
		\includegraphics[width=\textwidth]{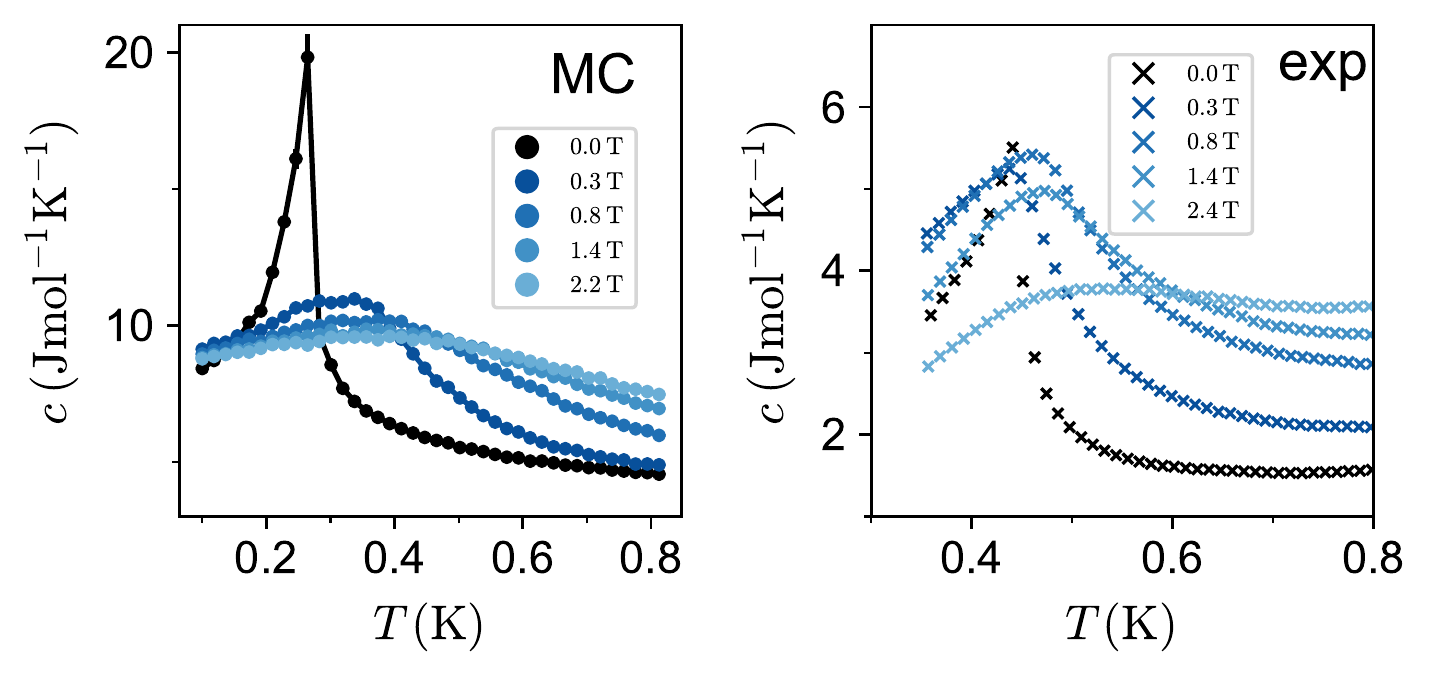}
	\end{minipage}
	\begin{minipage}{0.33\textwidth}
		\includegraphics[width=\textwidth]{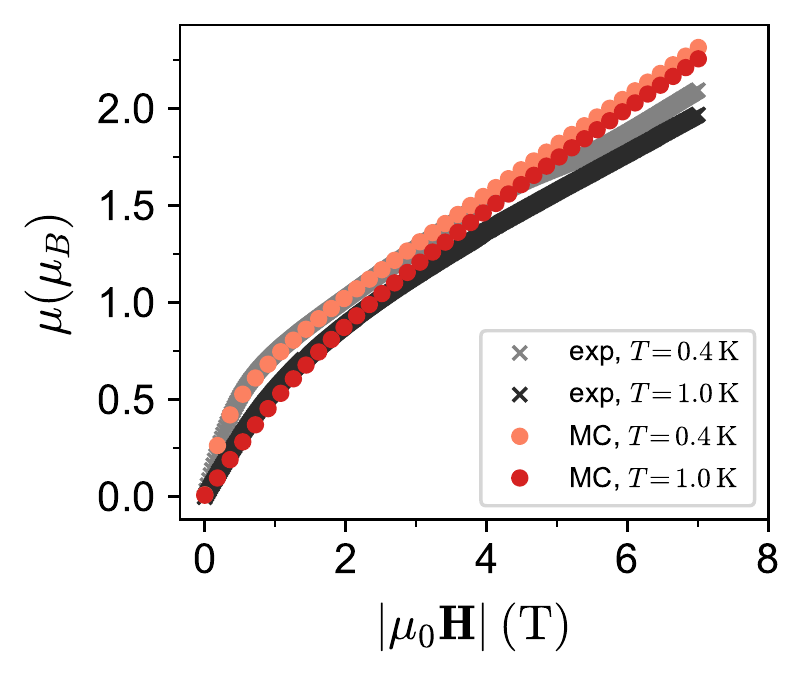}
	\end{minipage}
	\begin{textblock}{1}(-0.2,-2.8)
		\textbf{a.}
	\end{textblock}
	\begin{textblock}{1}(3.8,-2.8)
		\textbf{b.}
	\end{textblock}
	\begin{textblock}{1}(7.9,-2.8)
		\textbf{c.}
	\end{textblock}
    \caption{\textbf{a.} Specific heat in an external field computed from MC simulations with dipolar interactions and system parameters $L = 12$, $R = 0.55$, field orientated in the $(100)$ direction. Results are similar for field oriented in the $(110)$ direction. 
    \textbf{b.} Specific heat measured in experiments on a powder sample for external fields over the same temperature range.
    Simulations and experiments show the same qualitative picture as the external field is increased, with the peak broadening and shifting to higher temperature as well as a shift of entropy to higher temperatures.
    \textbf{c.} Magnetization in an external field at low temperatures, comparing simulations with dipolar interactions ($L = 12, R = 0.55$, field in $(100)$ direction) and experiments. We have verified that the simulation results do not depend strongly on field orientation. 
    Experiment and simulations are in agreement for $\abs{\mu_0 \mathbf{H}} \lesssim 3 \: \mathrm{T}$.
	}
\label{fig:dipolar_exp}
\end{figure}

We have verified that the onset of order at low temperature does not affect the properties of the model for temperatures $T > 1 \: \mathrm{K}$; comparing the susceptibility, magnetization in an external field and spin structure factor with and without dipolar interactions finds only very small differences. 
Therefore we conclude that the inclusion of dipolar interactions does not destroy the finite temperature spin liquid regime $1 < T < 4 \: \mathrm{K}$. 
Below this temperature we would expect the model specified by eq. \ref{eq:dip_ham} to reproduce (at least some) experimental properties.
First of all, there is quantitative discrepancy in ordering temperatures between our simulations and experiment, however this is relatively common in classical systems, see eg. ref. \cite{Yan2017}.
Furthermore, the inclusion of dipolar interactions qualitatively reproduces the experimentally observed behaviour of the specific heat in an external field, as shown in figs. \ref{fig:dipolar_exp}a,b.
Since classical models are not able to capture the bump in specific heat observed in experiments around $4 \: \mathrm{K}$, there will be less entropy released, which at least partially explains the much higher values of the specific heat in our classical simulations.
Furthermore, the model with dipolar interactions is able to reproduce the magnetization in an external field up to about $3 \: \mathrm{T}$, as shown in fig. \ref{fig:dipolar_exp}c.

Therefore with the inclusion of dipolar interactions the extended CPy Hamiltonian is able to account for 
experimental features in the regime $T_c = 0.43 < T < 1 \: \mathrm{K}$, $0 \leq \abs{\mu_0 \mathbf{H}} \lesssim 3 \: \mathrm{T}$. 
Obtaining a closer fit between experiment and theory here would probably require the inclusion of quantum effects in large-scale systems and/or analysing the effect of further (small) perturbations to eq. \ref{eq:dip_ham} 
as well as analysis of the effect of powder samples in the case of measurements in a finite external field.
\clearpage
\bibliography{supp_refs}